\documentclass[sn-mathphys,Numbered]{sn-jnl}

\usepackage{graphicx}%
\usepackage{multirow}%
\usepackage{amsmath,amssymb,amsfonts}%
\usepackage{amsthm}%
\usepackage{mathrsfs}%
\usepackage[title]{appendix}%
\usepackage{xcolor}%
\usepackage{textcomp}%
\usepackage{manyfoot}%
\usepackage{booktabs}%
\usepackage{algorithm}%
\usepackage{algorithmicx}%
\usepackage{algpseudocode}%
\usepackage{listings}%
\usepackage{caption}%
\usepackage{cleveref}%
\usepackage{subcaption}%
\usepackage{placeins}

\theoremstyle{thmstyleone}%

\theoremstyle{thmstyletwo}%

\theoremstyle{thmstylethree}%

\crefname{figure}{Figure}{Figures}
\crefname{section}{Section}{Sections}
\crefname{equation}{Equation}{Equations}

\begin{document}

\title[Article Title]{Principal Component Analysis and K-Means Clustering of Fuel-Air Mixing in Gas Turbine Combustors}

\author*[1, 2]{\fnm{David} \sur{Salvador-Jasin}}\email{dsalvadorjasin@turing.ac.uk}

\author[1]{\fnm{A Duncan} \sur{Walker}}\email{a.d.walker@lboro.ac.uk}

\author[1]{\fnm{Jon F} \sur{Carrotte}}\email{j.f.carrotte@lboro.ac.uk}

\affil[1]{\orgdiv{National Centre for Combustion and Aerothermal Technology}, \orgname{Loughborough University}, \orgaddress{\city{Loughborough}, \postcode{LE11 3GR}, \country{UK}}}

\affil[2]{\orgdiv{Research Engineering Group}, \orgname{The Alan Turing Institute}, \orgaddress{\street{British Library}, \city{London}, \postcode{NW1 2DB}, \country{UK}}}

\abstract{As a direct consequence of liquid kerosene injection, aeroengine combustors may be categorized as non-premixed combustion systems, characterized by a swirl-stabilized and highly complex flow field. In addition to the flow of air through the fuel injector, there are a large number of other features through which oxidizer can enter the heat release region. These can have an impact on local fuel-air mixing, inducing strong spatial and temporal variations in stoichiometry, thereby affecting emissions and combustion system performance. This paper discusses a novel statistical methodology, based on Principal Component Analysis (PCA) and K-means clustering, that aims to improve understanding of fuel-air mixing in realistic aeroengine combustors. The method is applied in a postprocessing step to data sampled from a Large Eddy Simulation (LES), where every chamber inflow has been tagged with a unique passive scalar, which allows it to be traced across space and time. PCA is used to construct a low-dimensional, visually interpretable representation of a spatially localized fuel-air mixing process, while K-means clustering is employed to produce an unsupervised discretization of the flow field into regions of similar fuel-air mixing characteristics. The proposed methodology is computationally inexpensive, and the easily interpretable outputs can help the combustion engineer make better informed decisions about combustor design.}

\keywords{aeroengine, combustion, LES, PCA, clustering, K-means, low-order modeling, unsupervised machine learning}

\maketitle

\section{Introduction}\label{sec:intro}

The increasingly high pressure ratios that are required in modern civil aeroengines to improve their thermodynamic efficiency result in ever higher combustor inlet pressures and temperatures. This presents a major challenge for the implementation of premixed ultra-low-emission systems due to the risk of autoignition and flashback \cite{Lefebvre2010}. For this reason, aeroengine combustors typically operate with direct fuel injection. In other words, aviation kerosene is injected into the combustion chamber in liquid form, where it evaporates and mixes with the air prior to combustion in a diffusion flame. Achieving rapid evaporation and fuel-air mixing is essential for clean and efficient combustion \cite{Lefebvre2010}. The fuel injector assembly plays a crucial role in this process. In modern fuel injectors, a thin and slowly moving fuel film is spread over a surface known as the prefilmer, where the momentum associated with the large amounts of air from the upstream compressor is used to break the fuel film into a finely atomized spray which can evaporate quickly \cite{Lefebvre2010, Lefebvre2017}. In order to enhance the mixing between the fresh reactants and the burnt products, which provide a continuous source of ignition to the fresh mixture, the fuel injector generates high swirl and causes a large region of recirculating flow with high levels of unsteadiness and turbulence, known as the central recirculation zone (CRZ) \cite{Lefebvre2010}. The aerodynamic blockage caused by the CRZ anchors the flame and results in what is known as a swirl-stabilized flame burning in diffusion or non-premixed mode.

\begin{figure}[h]
    \centering
    \includegraphics[height=6cm]{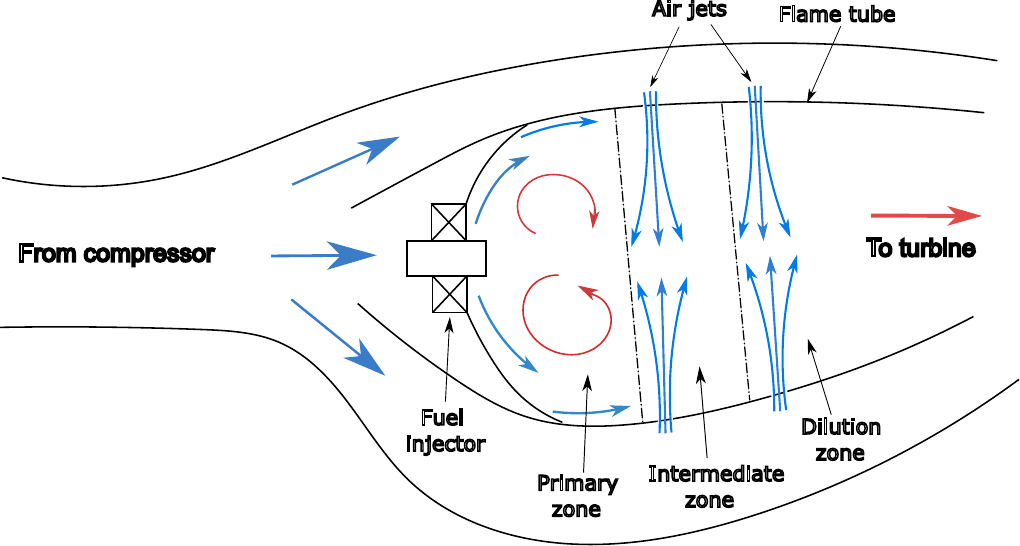}
    \captionsetup{skip=10pt, justification=raggedright}
    \caption{Diagram of the airflow pattern inside a modern aeroengine combustor. Blue arrows represent cold streams, while red arrows represent hot streams.}
    \label{fig:combustor-diagram}
\end{figure}

\cref{fig:combustor-diagram} illustrates the airflow pattern inside a modern rich-burn aeroengine combustor. They are of the annular type: a single flame tube, in annular shape, contained in a pressure casing, with an array of fuel injectors uniformly distributed in the circumferential direction. The chamber is open at the front to the compressor and at the rear to the turbine. Since not all of the air exiting the compressor is required for combustion, it is introduced into the chamber in a staged fashion. As a result, the combustor is split into separate zones with different overall fuel-to-air ratios (FAR). While this depends on the combustor architecture, generally three different zones can be distinguished: the primary, intermediate, and dilution zones \cite{Lefebvre2010, Rolls-Royce2015}, as illustrated in \cref{fig:combustor-diagram}. The objective of the primary zone is to anchor the flame and achieve essentially complete combustion. The majority of the air in the primary zone is introduced through the fuel injector assembly. The bulk of the additional air in the other zones is injected through holes in the flame tube, in the form of jets whose penetration must be optimized for improved mixedness with the main stream. The role of the intermediate zone is to encourage the oxidation of incomplete products of combustion such as carbon monoxide, unburned hydrocarbons and soot particles. Finally, the purpose of the dilution zone is to admit the air remaining after combustion and wall cooling, and to obtain an outlet stream with a temperature distribution that is acceptable to the downstream turbine.

The combustor may therefore be viewed as a mixing chamber in which the stoichiometry must be tightly controlled in the different regions of the flame tube. Good mixing quality, and in the correct proportions, between the multiple air inflows and the fuel stream is essential for high combustion efficiency and reduced pollutant formation. However, within each region, the injection of liquid fuel together with the large number of features through which the oxidizer can enter the chamber means that the degree of mixedness between the fuel and air streams can vary significantly from one location to another. In addition to the spatial variation in mixing, the high levels of flow unsteadiness and turbulence will result in strong temporal fluctuations in the local FAR. Brend et al. \cite{Brend2020} showed, using volumetric particle image velocimetry measurements, that the jet flows influence each others' behavior and that their unsteady interaction is driven by the boundary conditions imposed by the external annular passages. In an earlier experiment, Hughes and Carrotte \cite{Hughes2004} showed that the unsteadiness of the external flow approaching a jet port, where length scales much larger than the port diameter could be identified, influenced the time-dependent flow field within the flame tube. The unsteady interaction between the different inflows and the large range of length scales within the chamber will affect fuel-air mixing and emissions performance, such as the formation of soot and nitrogen oxides \cite{Lefebvre2010}.

In order to better understand the role of each inflow on the local fuel-air mixing, a new statistical methodology, based on Principal Component Analysis (PCA) and K-means clustering, is proposed. The main idea behind PCA is to summarize the correlation structure of a multivariate data set using a reduced number of dimensions, while retaining as much variance/energy as possible \cite{Jolliffe2002}. As will be shown in later sections, PCA can provide an easily interpretable summary of a complex local mixing process characterized by a large number of inflows simultaneously influencing the local FAR. PCA has gained popularity in recent years in the field of turbulent combustion, due to its ability to identify low-dimensional thermochemical manifolds in turbulent reacting systems. In an early publication, Parente et al. \cite{Parente2009} discussed the application of PCA to measurement data of different flames and pointed out that the number of modes required to accurately reconstruct the state variables strongly depended on the physical processes affecting the flame (e.g. extinction). In a more recent study, D'Alessio et al. \cite{DAlessio2020} applied a local formulation of PCA to simulation data of a turbulent reacting jet, identifying clusters of spatial locations with similar thermochemical states in an unsupervised fashion. PCA is in fact a more generic name for the well-known Proper Orthogonal Decomposition (POD) in fluid dynamics \cite{Berkooz1993}. In its classical implementation, the POD is applied to a matrix of turbulent velocity signals sampled from a spatial grid, with the aim of identifying coherent flow structures of high turbulent kinetic energy. PCA applies exactly the same mathematical operations as the POD, but the name PCA is here preferred because the name POD is typically reserved for dynamical systems, such as velocity or vorticity spatio-temporal fields \cite{Brunton2019}. While PCA can be used to simplify the analysis of a local fuel-air mixing process, K-means clustering can be employed to find spatial locations across the combustor where the fuel-air mixing trends are similar, thereby generating a reduced number of spatial clusters in an unsupervised fashion.

\section{Numerical Methodology}\label{sec:numerical-methodology}

\subsection{Large Eddy Simulation}\label{sec:large-eddy-simulation}
It is not currently possible to obtain the datasets required for the proposed statistical methodology experimentally, and we therefore rely on numerical modeling instead. It is beyond the scope of this work to provide a detailed discussion on the application Computational Fluid Dynamics (CFD) for modeling gas turbine combustion systems \cite{Poinsot, Chen2020, Veynante2002, Vervisch2015}. In short, the prohibitive computational cost of Direct Numerical Simulation (DNS) (where all turbulence scales, down to the Kolmogorov scale, are explicitly solved) makes it unfeasible to model realistic combustor geometries. On the other hand, it is well known that Reynolds-Averaged Navier Stokes (RANS) turbulence models are generally not well suited for strongly swirling flows such as those found in aeroengine combustors, and the solutions correspond to averages over time. However, Large Eddy Simulation (LES) explicitly solves the system-dependent large scales while only modeling the self-similar small eddies, which lend themselves to simpler and more universal models \cite{Pope2000}, and therefore the prediction error is reduced compared to RANS. More importantly, the main aim of this work is to perform a statistical analysis of the multivariate correlations between passive scalars representing the time-history of the fuel-air mixing process, which are not possible to obtain with RANS. In view of this, the considerably larger computational cost of LES over RANS is justified and LES has been chosen as the preferred method for capturing the complex mixing inside the aeroengine combustor.

In order to investigate fuel-air mixing under realistic conditions, reacting flow large eddy simulations were performed, where combustion was modeled using the Flamelet Generated Manifold (FGM) approach \cite{Fiorina2015, Chen2020}. The liquid fuel spray was modeled using a two-way coupled Eulerian-Lagrangian framework. Primary atomization was neglected; a spray consisting of a distribution of droplet sizes was injected directly into the flow field just downstream of the fuel prefilmer. As the droplets travel downstream, they undergo secondary atomization and evaporation \cite{Lefebvre2017}. This is a typical approach in LES of industrially relevant gas turbines due to the computational cost and complexity of combustion and spray modeling \cite{Chen2020}.

\subsection{Investigated Geometries}\label{sec:investigated-geometries}
\begin{figure}[h]
\begin{subfigure}{.4\textwidth}
	\centering
	\includegraphics[trim={1.5cm, 0cm, 7cm, 1.1cm}, clip, width=\linewidth, height=5cm, keepaspectratio]{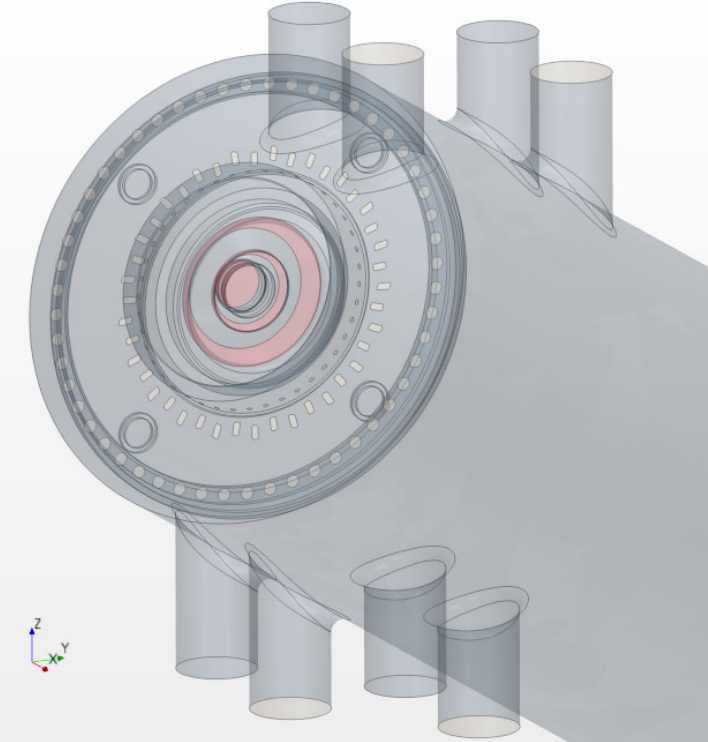}
	\caption{Computational model}
\end{subfigure}
\hspace*{.10\textwidth} 
\begin{subfigure}{.5\textwidth}
	\centering
	\includegraphics[trim={1.5cm, 0.2cm, 6.5cm, 3cm}, clip, width=\linewidth, height=5cm, keepaspectratio]{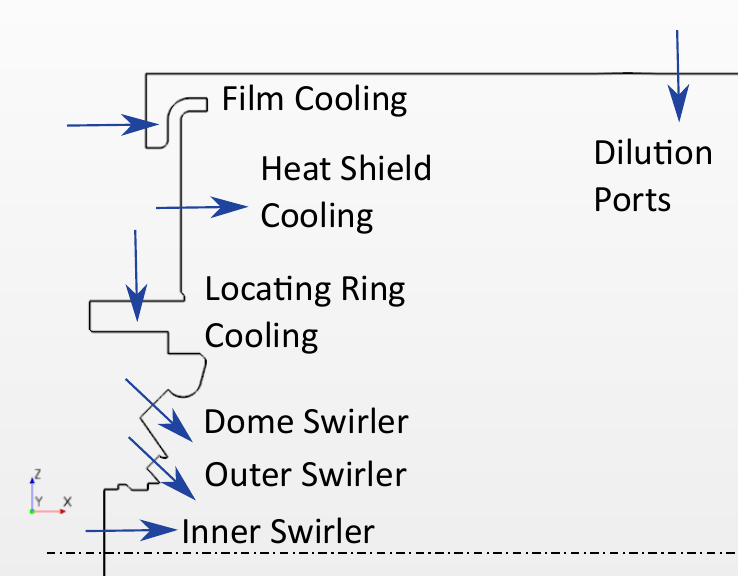}
    \caption{Cut-through of the upper half}
\end{subfigure}

\caption{Computational model of the experimental test rig with indication of all inlet air streams. The geometry is circumferentially symmetrical about the injector centerline, indicated by the dashed-dotted line. Dilution ports can be found on each side of the rig along the $z$-axis.}
\label{fig:test-rig}
\end{figure}

Reacting-flow large eddy simulations of two different models were performed, and the proposed statistical methodology was applied to the sampled data in a post-processing step. The first geometry corresponds to a model of an experimental test rig designed to measure pollutant emissions at low to intermediate pressures. It consists of a single fuel injector inside a cylindrical combustion chamber which is symmetrical about the injector centerline. This architecture results in a simpler flow field compared to a gas turbine combustor, but it is sufficiently representative of engine conditions to provide a valuable dataset to test the proposed statistical methodology. The computational model of the experimental test rig is illustrated in \cref{fig:test-rig}.

\begin{figure}[h]
\begin{subfigure}{.45\textwidth}
	\centering
	\includegraphics[trim={10cm, 1.5cm, 6cm, 1.5cm}, clip, width=1.1\linewidth]{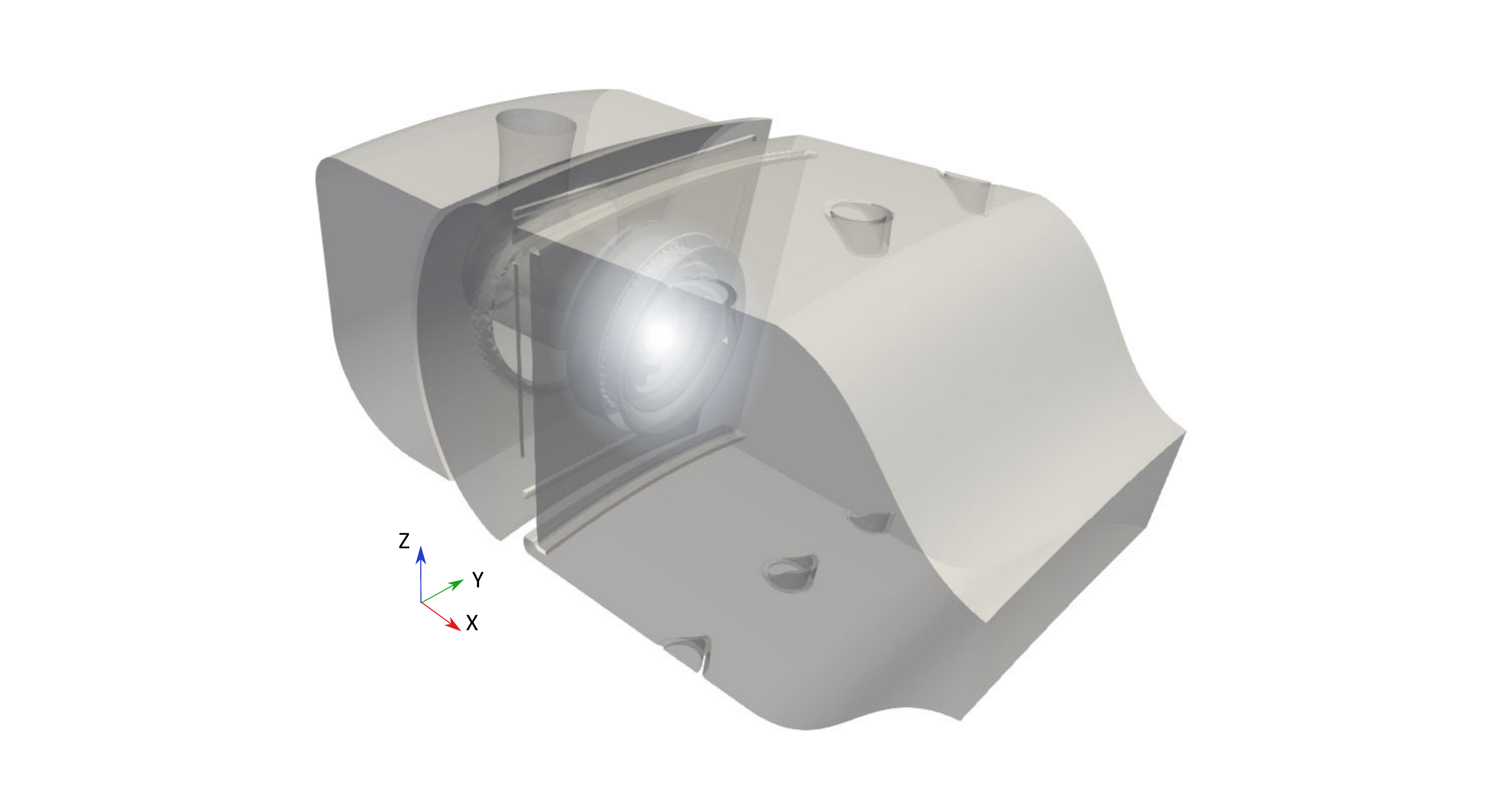}
	\caption{Computational model}
\end{subfigure}
\begin{subfigure}{.55\textwidth}
	\centering
	\includegraphics[trim={6cm, 2.3cm, 11cm, 2.3cm}, clip, width=1.1\linewidth, height=0.75\linewidth]{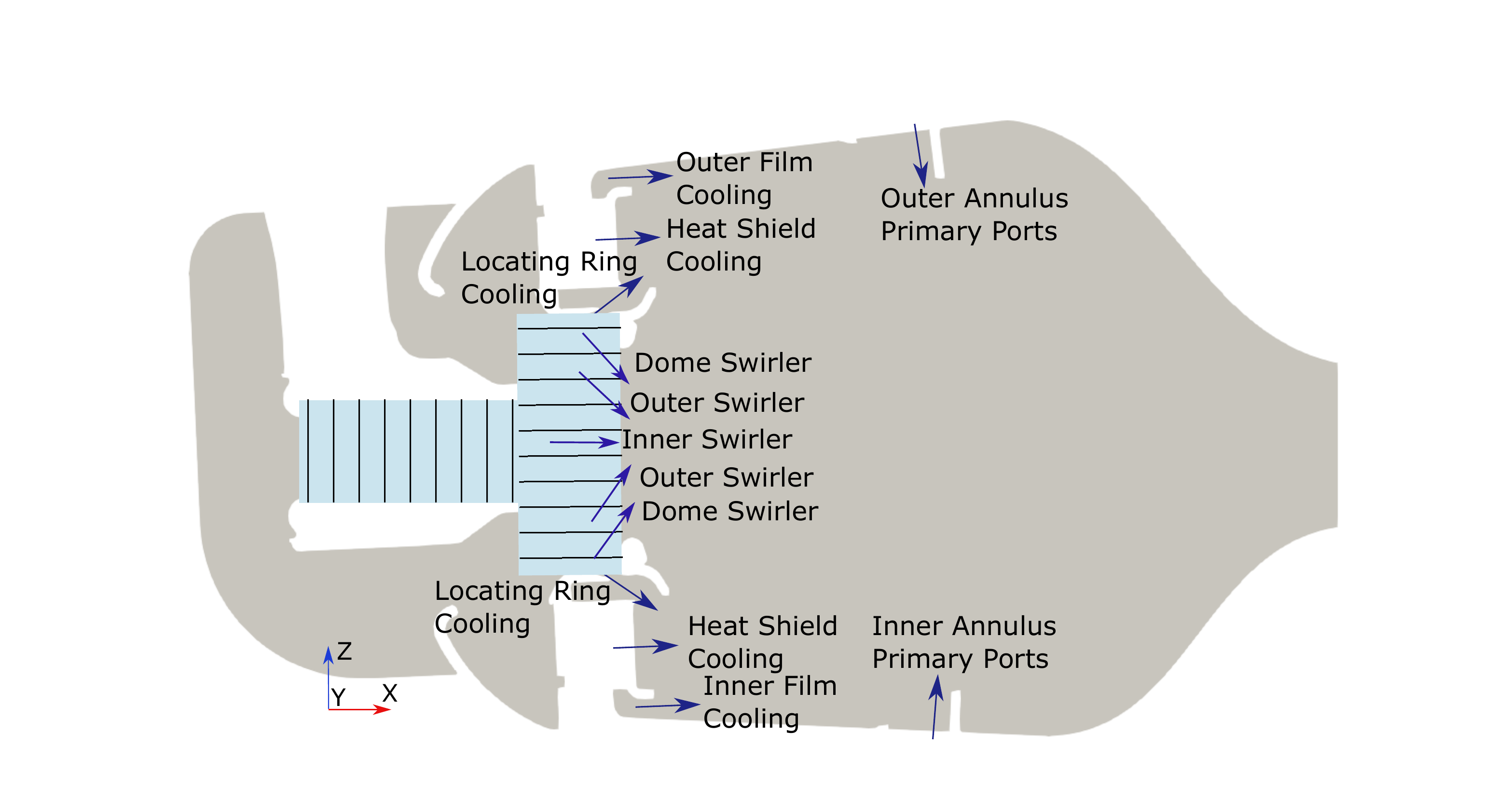}
	\caption{Cut-through of the domain}
\end{subfigure}
\caption{The detailed single sector annular combustor with indication of all inlet air streams. Periodic boundary conditions were imposed on the side walls. The internal components of the fuel injector have been concealed due to commercial confidentiality.}
\label{fig:detailed-combustor}
\end{figure}

The second geometry corresponds to a detailed model of a single sector of an annular combustion chamber operated at take-off (high power) conditions, and is therefore more representative of engine conditions than the experimental test rig. The computational model is illustrated in \cref{fig:detailed-combustor}. Periodic boundary conditions were applied on the side walls of the single sector to simulate the presence of adjacent fuel injectors.

Note that in both cases, only the primary combustion zone (i.e. the upstream region of the combustor, up to the first row of wall air jets - see \cref{fig:combustor-diagram}) was modeled in the large eddy simulations. The reason for this modeling approach is that one of the purposes of these simulations was to investigate soot formation for different fuel injector configurations, which mainly takes place in the primary combustion zone under rich-burn conditions \cite{Lefebvre2010}.

\FloatBarrier

\subsection{Passive Scalars for Stream Tracing}\label{sec:passive-scalars}
\begin{figure}[h]
	\begin{subfigure}{.5\textwidth}
		\centering
		\includegraphics[trim={0.2cm, 2.2cm, 0.2cm, 7.2cm}, clip, width=0.85\linewidth]{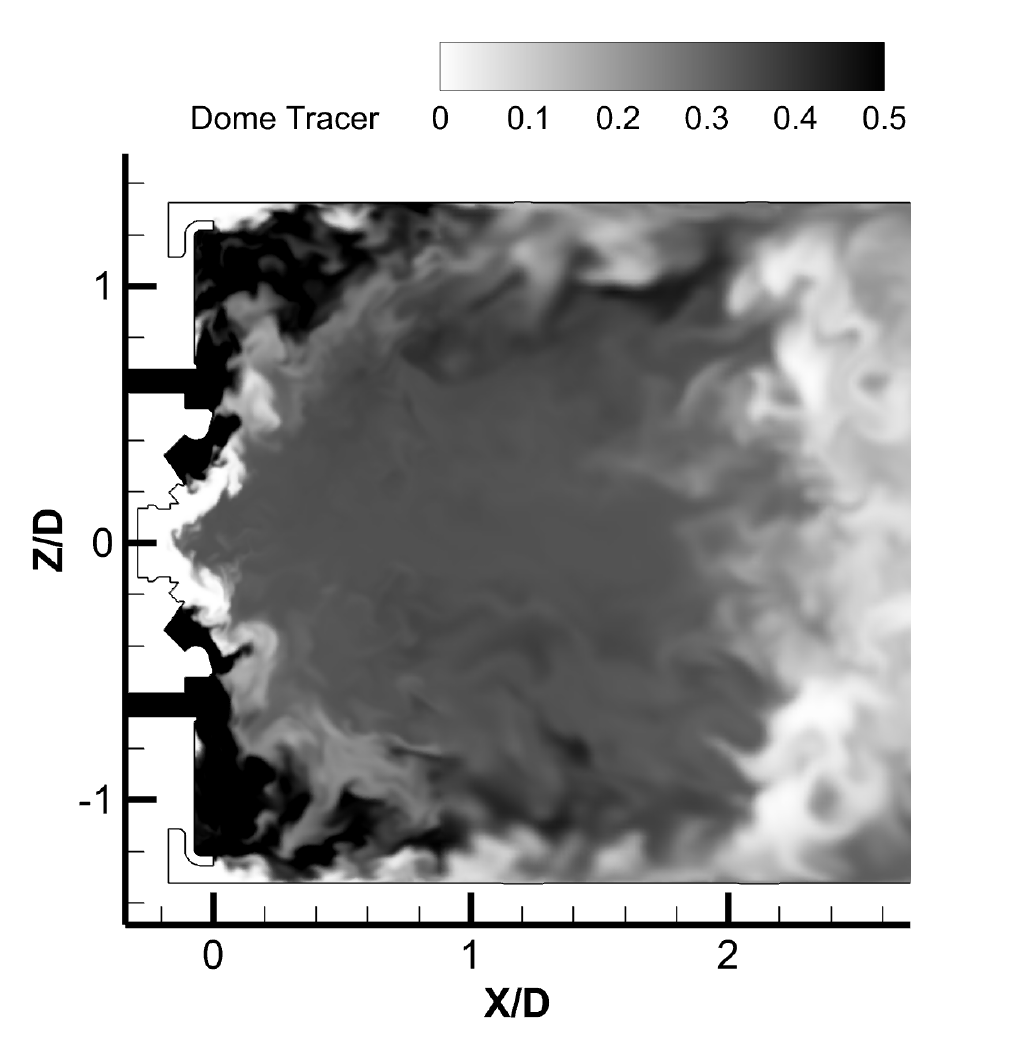}
		\caption{Instantaneous dome swirler tracer}
	\end{subfigure}
	\begin{subfigure}{.5\textwidth}
		\centering
		\includegraphics[trim={0.2cm, 2.2cm, 0.2cm, 7.2cm}, clip, width=0.85\linewidth]{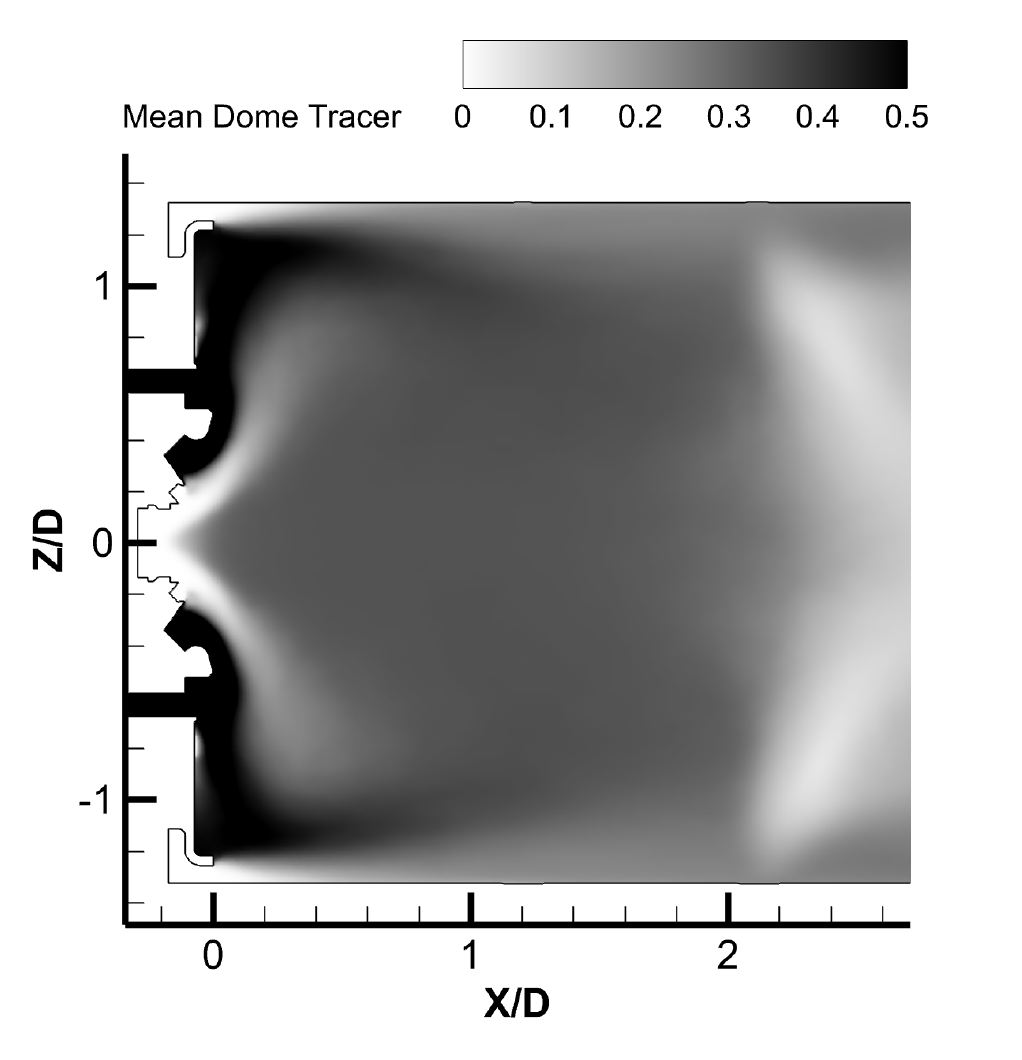}
		\caption{Time-averaged dome swirler tracer}
	\end{subfigure}
	\newline
	\begin{subfigure}{.5\textwidth}
		\centering
		\includegraphics[trim={0.2cm, 2.2cm, 0.2cm, 6.2cm}, clip, width=0.85\linewidth]{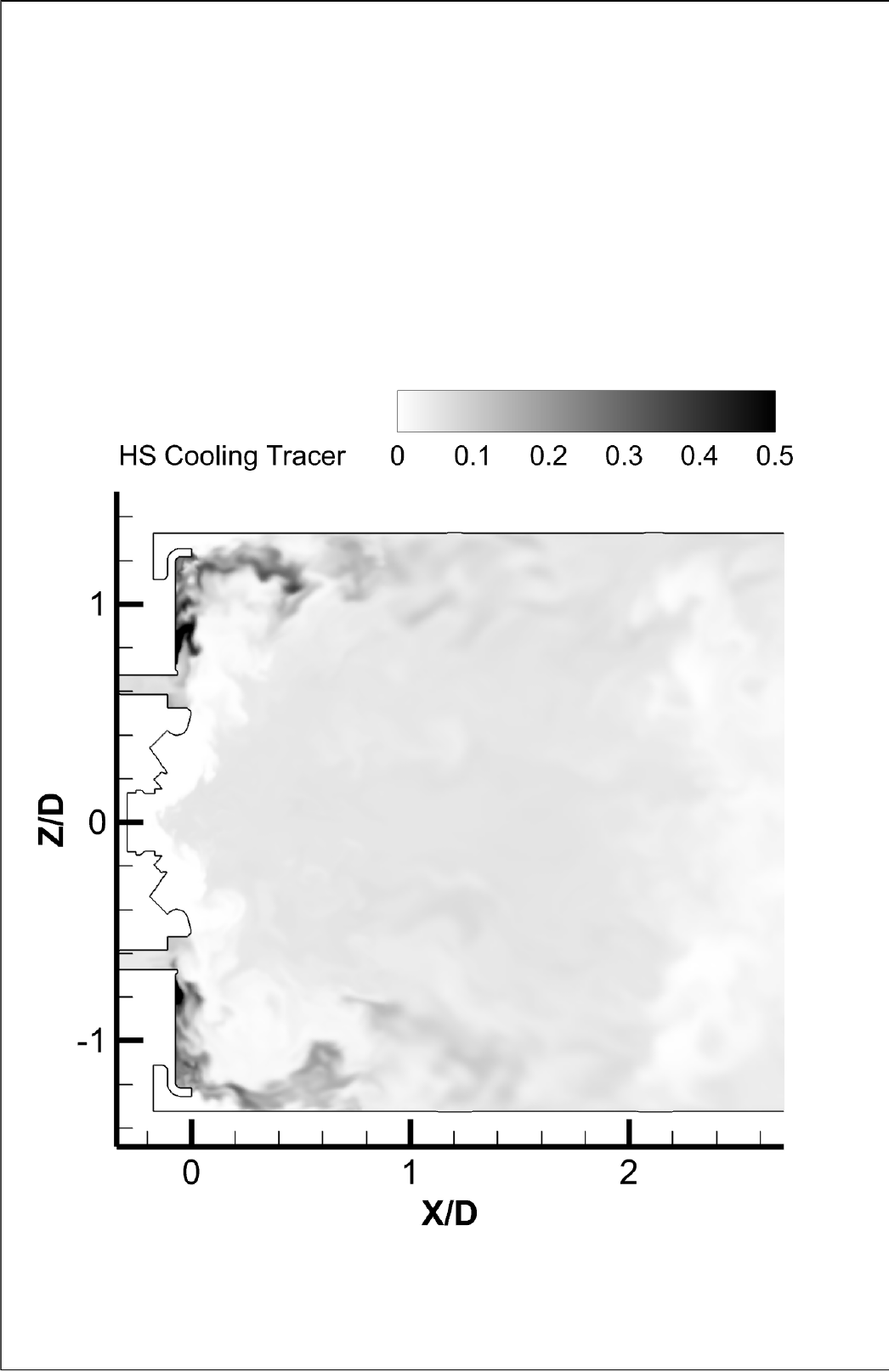}
		\caption{Instantaneous heat shield cooling tracer}
	\end{subfigure}
	\begin{subfigure}{.5\textwidth}
		\centering
		\includegraphics[trim={0.2cm, 2.2cm, 0.2cm, 6.2cm}, clip, width=0.85\linewidth]{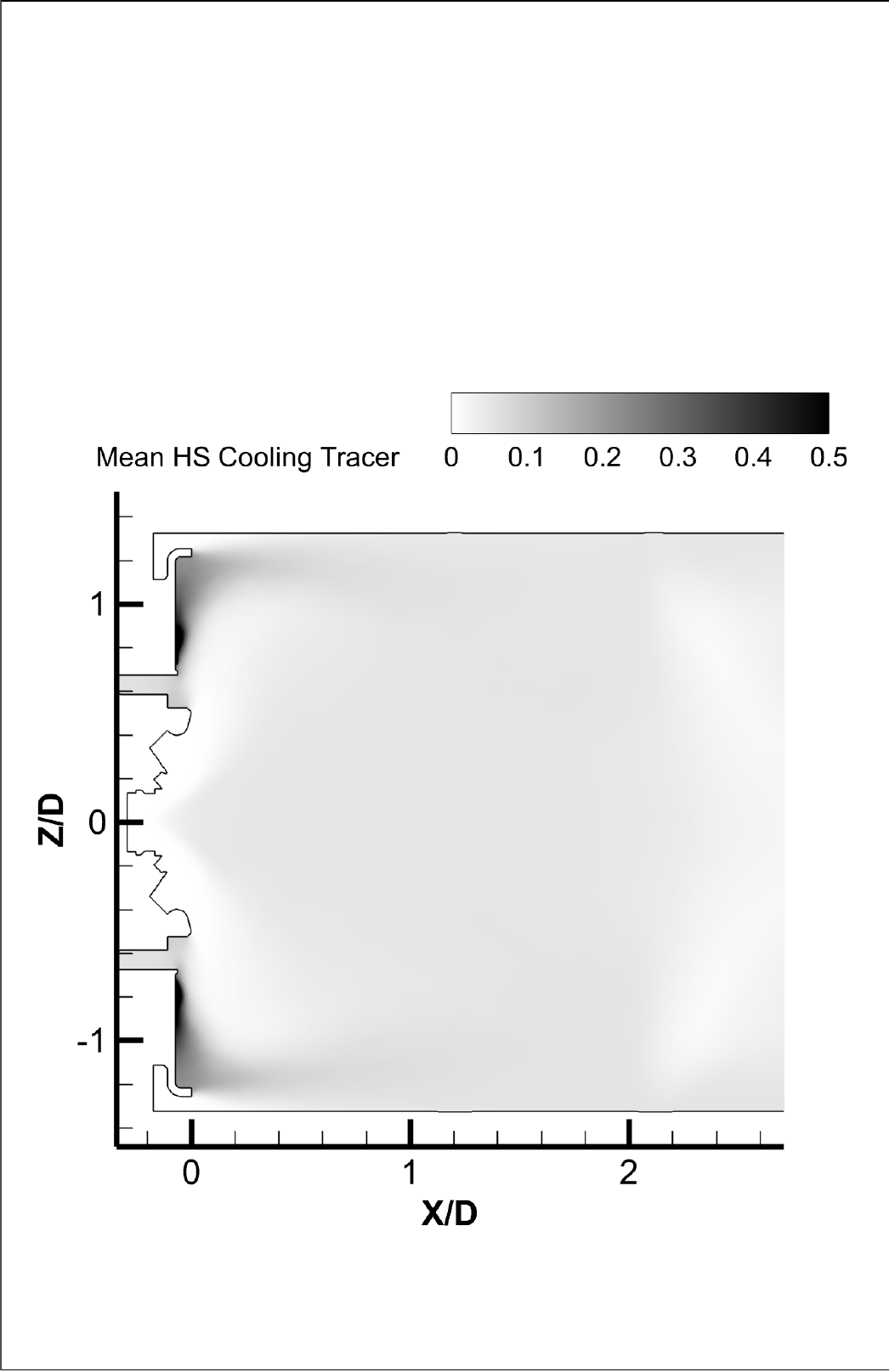}
		\caption{Time-averaged heat shield cooling tracer}
	\end{subfigure}

\caption{Instantaneous and time-averaged profiles of the dome swirler and heat shield cooling passive scalars across the mid-plane of the experimental test rig.
The coordinates have been normalized by one injector diameter.}
\label{fig:passive-scalar-distributions}
\end{figure}

In order trace the different inflows into the chamber across space and time, new variables, known as passive scalars, can be defined. They are referred to as passive scalars because they have no effect on the flow field. They are the numerical equivalent of a tracer dye in a fluids experiment. Each passive scalar must be solved by an additional transport equation in the LES calculation. In flamelet-based combustion modeling, the passive scalar that tracks the fuel stream in a non-premixed flame is known as the mixture fraction ($\xi$) \cite{Peters2000, Veynante2002}. At any location in the combustion chamber, the mixture fraction indicates the local ratio of the gaseous mass flux ($\dot{m}$) originating from the fuel feed to the total mass flux:

\begin{equation}
    \xi=\frac{\dot{m}_1}{\dot{m}_1 + \dot{m}_2},
    \label{eq:mix-frac}
\end{equation}

\noindent where subscripts 1 and 2 refer to the fuel and oxidizer feeds, respectively. The boundary conditions for the mixture fraction are $\xi=1$ in the fuel feed and $\xi=0$ in the oxidizer feed. Analogous passive scalars can be defined for each individual oxidizer stream into the chamber (cooling flows, port flows, fuel injector air flows, etc.). If each stream is tagged with a unique passive scalar and is assigned a value of unity at its corresponding inlet and zero at every other inlet, the following equation holds at every location in the combustion chamber at any point in time:

\begin{equation}
    \xi + \sum_{j=1}^P Y_j = 1,
\end{equation}

\noindent where $Y_j$ is the passive scalar tagging the $j^{th}$ oxidizer stream and $P$ is the total number of oxidizer passive scalars. Therefore, like the mixture fraction, $Y_j$ indicates the local fraction of mass flux originating from the $j^{th}$ oxidizer inlet (see \cref{fig:passive-scalar-distributions} for an illustration). As was already discussed in \cref{sec:intro}, the instantaneous local FAR, which is directly related to $\xi$, is a parameter of key importance for high combustion efficiency and low pollutant formation. Since the instantaneous stoichiometry anywhere in the combustor is governed by the degree of mixedness between the different oxidizer streams and the fuel stream, passive scalars are a useful tool as they enable the influence of individual streams on the mixing to be identified.

\section{Statistical Methodology}\label{sec:statistical-methodology}

\subsection{Covariance and Correlation}\label{sec:covariance-correlation}
If $u$ and $v$ are two random variables, the covariance between them is defined as:

\begin{equation}
    \sigma_{u,v}^{2} = \lim_{N \to \infty} \frac{1}{N} \displaystyle \sum_{i=1}^{N} [(u_i - \overline{u}) \, (v_i - \overline{v})],
    \label{eq:covariance}
\end{equation}

\noindent where $\overline{u}$ and $\overline{v}$ are the mean or expected values of $u$ and $v$, respectively, and $N$ is the number of observations or snapshots sampled. The covariance is a measure of the joint variability of two random variables, and its sign is indicative of the linear relationship between them. A positive covariance indicates that $u$ and $v$ tend to increase or decrease together, while a negative covariance indicates that when one increases, the other tends to decrease. The magnitude of the covariance is not easily interpretable and it depends on the magnitude of the variables. For this reason it is typically normalized, giving rise to the linear correlation coefficient:

\begin{equation}
    \rho_{u,v} = \frac{\sigma_{u,v}^{2}}{\sigma_u \sigma_v},
    \label{eq:pearson-correlation}
\end{equation}

\noindent where $\sigma_{u}$ and $\sigma_{v}$ are the standard deviations of $u$ and $v$, respectively. Unlike the covariance, the correlation coefficient is bounded between -1 and 1, where a value of 1 indicates perfect linear correlation, a value of -1 indicates perfect linear anti-correlation and a value of 0 means that no linear correlation exists.

Covariance and correlation are useful concepts to determine the degree of mixing between two streams at a given location in the combustion chamber. The idea was previously used by Giusti et al. \cite{Giusti2019} to analyze flow inhomogeneities and mixture fraction fluctuations at a combustor outlet using LES data. A high positive correlation over time between two stream tracers (e.g., the inner swirler passive scalar and the mixture fraction) at a fixed location in a non-premixed combustion system suggests that their mixing is nearly complete by the time they reach this point. On the other hand, a high negative correlation between a different oxidizer tracer (e.g. a wall cooling stream) and the mixture fraction at this same location indicates that the oxidizer in question has barely mixed with the fuel at upstream locations.

\subsection{Principal Component Analysis}\label{sec:pca}
Analyzing correlation coefficients to understand if two streams have mixed or are in the process of mixing is manageable in a simple system consisting of only a few streams. However, if we are interested in analyzing the mixing between all the streams in a realistic combustion chamber, which could consist of many inflows (e.g. see \cref{fig:detailed-combustor}), it is helpful to reduce the dimensionality of the problem for ease of interpretation. As discussed in \cref{sec:intro}, PCA is a mathematical technique for reducing the dimensionality of a multivariate dataset while retaining as much variability/variance as possible \cite{Jolliffe2002}. The idea behind PCA is that, although datasets relevant to engineering can have very high dimensionality, usually they are low-rank: there are a few dominant patterns that explain the high dimensionality \cite{Brunton2019}.

PCA is typically performed by applying the Singular Value Decomposition (SVD) to a data matrix $\mathbf{X}$ of dimensions $(N \times P)$, where $N$ is the number of observations or snapshots and $P$ is the number of variables \cite{Jolliffe2002, Brunton2019}. In the application proposed in this work, $P$ represents the total number of passive scalars tagging the different inflows into the combustion chamber, and $N$ represents the number of temporal observations sampled from the LES calculation. The result of SVD is a new set of orthogonal modes called principal components (PCs) that successively maximize the variance of the original data matrix $\mathbf{X}$. This is identical to the procedure followed in the POD in fluid dynamics, where $N$ represents the number of (typically) velocity observations over time, and $P$ represents the number of points in a spatial grid. In POD, principal components are usually referred to as spatial modes \cite{Berkooz1993}, and they represent regions of the flow field with correlated flow motions of high turbulent kinetic energy.

In order to understand the relationship between the concept of covariance discussed in \cref{sec:covariance-correlation} and PCA, it is useful to start by performing the eigenvalue decomposition of the covariance matrix of $\mathbf{X}$:

\begin{equation}
    \mathbf{S} = \frac{1}{N} \mathbf{X}^T \mathbf{X} = \mathbf{A L A}^T,
    \label{eq:covariance-eigenvalue-decomposition}
\end{equation}

\noindent where $\mathbf{S}$ is the $(P \times P)$ covariance matrix of $\mathbf{X}$ (previously mean-centered), whose entries correspond to the covariances defined by \cref{eq:covariance}, $\mathbf{A}$ is the $(P \times P)$ matrix of eigenvectors $\mathbf{a}_j$, with $j \in \{1, 2, 3, ..., P\}$, and $\mathbf{L}$ is the $(P \times P)$ diagonal matrix of corresponding eigenvalues $l_j$. The eigenvectors $\mathbf{a}_j$ are in fact the PCs of $\mathbf{X}$, and their corresponding eigenvalue $l_j$ is the variance of $\mathbf{X}$ explained by $\mathbf{a}_j$. The PCs are perpendicular to each other and are sorted in descending order of explained variance.

\begin{figure}[h]
    \centering
    \includegraphics[height=5cm, trim={0 3cm 0 3cm},clip]{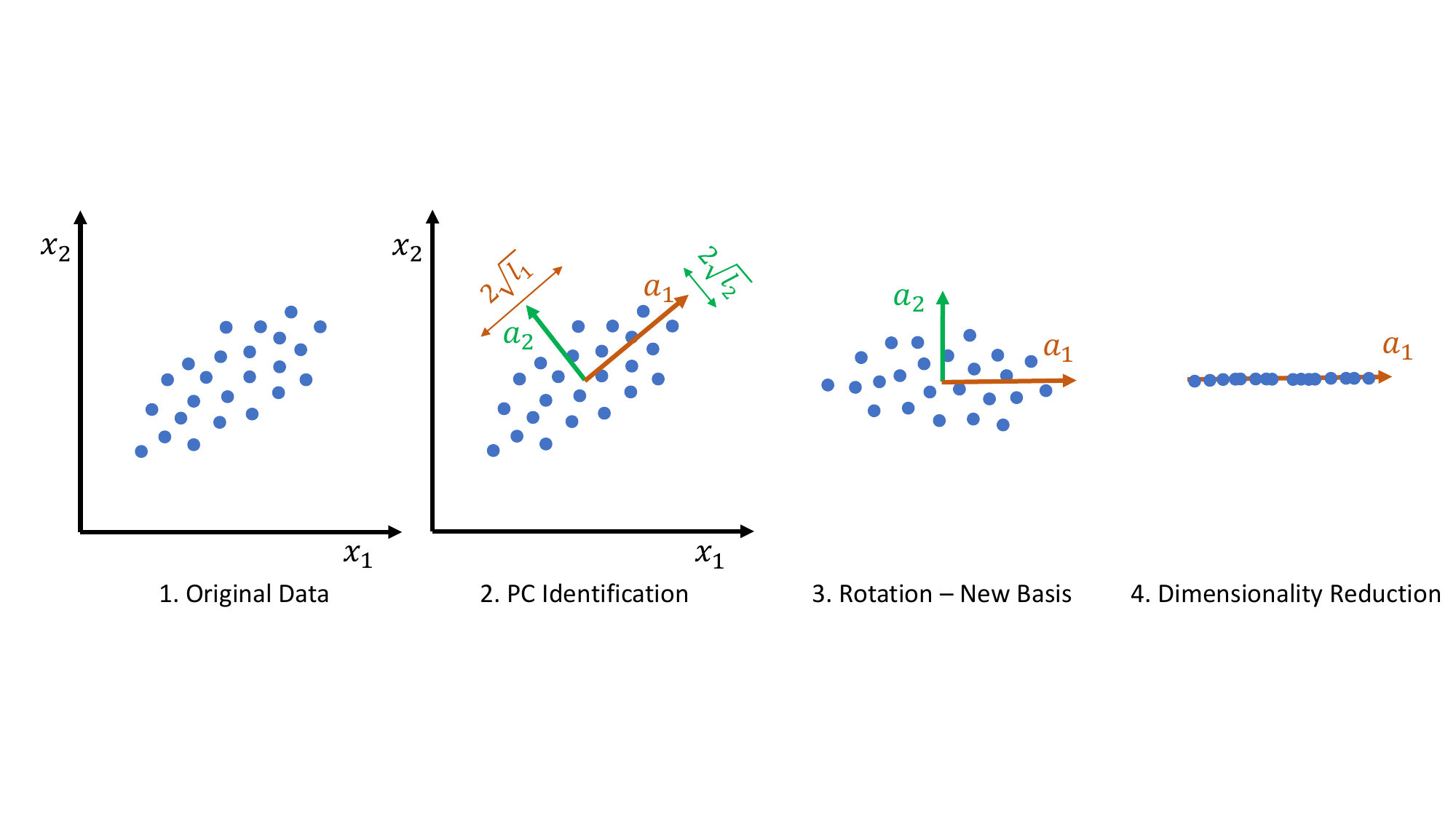}
    \caption{Illustration of dimensionality reduction with PCA. $\mathbf{a}_1$ and $\mathbf{a}_2$ are the first and second PCs, and $l_1$ and $l_2$ their corresponding variances.}
    \label{fig:pca-concept-illustration}
\end{figure}

The eigenvectors in $\mathbf{A}$ constitute the new orthogonal basis in which the original observations in $\mathbf{X}$ can be represented:

\begin{equation}
    \mathbf{Z} = \mathbf{X}\mathbf{A},
    \label{eq:pca-full-projection}
\end{equation}

\noindent where $\mathbf{Z}$ is a $(N \times P)$ matrix which is simply the orthogonal projection of the original dataset onto the new basis. The observations in $\mathbf{Z}$ are typically referred to as principal component scores \cite{Jolliffe2002}. Up to this point, the original variance of the dataset is conserved; it is simply redistributed across the newly defined dimensions (the PCs). Dimensionality reduction is achieved by projecting $\mathbf{X}$ onto a truncated basis:

\begin{equation}
    \mathbf{Z}_q = \mathbf{X}\mathbf{A}_q,
    \label{eq:pca-truncated-basis-projection}
\end{equation}

\noindent where $\mathbf{A}_q$ is a $(P \times Q)$ matrix containing the first $Q$ PCs, which explain $\sum_{j=1}^{Q} l_j$ of variance. If the dataset $\mathbf{X}$ is low-rank, as explained above, although the number of retained PCs ($Q$) may be much smaller than the total number of PCs ($P$), $\sum_{j=1}^{Q} l_j$ may represent a large fraction of the total variance of $\mathbf{X}$. Several strategies have been proposed in the literature to choose an adequate value of $Q$. A typical one is identifying the ``sharp edge" or ``elbow" in the eigenvalue distribution curve \cite{Brunton2019}. However, although crude, a commonly used technique is to truncate at a predetermined amount of total variance, e.g. 90\% or 95\%. The dimensionality reduction achieved through PCA is illustrated in \cref{fig:pca-concept-illustration} for a simple two-dimensional dataset. The two original dimensions, $\mathbf{x}_1$ and $\mathbf{x}_2$, are reduced to a single new dimension $\mathbf{a}_1$, which explains more variance than either $\mathbf{x}_1$ or $\mathbf{x}_2$ individually.

\subsection{Data Scaling}\label{sec:data-scaling}
Because PCA tries to capture the directions of maximum variance of the data matrix $\mathbf{X}$, the results are sensitive to the scales on which the variables are measured. Typically, PCA is said to be performed either on the covariance matrix, where the entries of $\mathbf{S}$ in \cref{eq:covariance-eigenvalue-decomposition} are computed from \cref{eq:covariance}, or on the correlation matrix, where the entries of $\mathbf{S}$ are calculated from \cref{eq:pearson-correlation} \cite{Jolliffe2002}. The choice depends on the application. In fluid dynamics, the POD is commonly performed on the covariance matrix because the aim is to find coherent flow structures of high turbulent kinetic energy (the variance of the velocity measurements). However, if the dataset is composed of variables of very different scales, such as a combustion manifold consisting of temperature and chemical species mass fractions, in order to capture the manifold's correlation structure while giving all variables equal importance, the right approach would be to perform PCA on the correlation matrix \cite{Parente2013}.

PCA on the covariance or the correlation matrix are only two options out of the possible range of data scaling strategies, and through an appropriate choice of scaling one can focus the analysis on specific variables of interest, as discussed by Parente and Sutherland \cite{Parente2013} in relation to turbulent combustion datasets. In the context of this investigation, the passive scalars represent fractions of mass flux originating from the different inlets into the chamber (see \cref{sec:passive-scalars}). It appears reasonable to assume that streams which represent larger fractions of the local mass flux (i.e. streams with larger passive scalar values at a given spatial location) will contribute more to the local fuel-air mixing. Performing PCA on the correlation matrix would effectively neglect this, as all streams would be given equal \textit{a priori} importance regardless of their passive scalar value.

Following the above discussion, the variables $\mathbf{x}_j$ in the data matrix $\mathbf{X}$ should be mean-centered and appropriately scaled before performing PCA:

\begin{equation}
    \widetilde{\mathbf{x}}_j = \frac{\mathbf{x}_j - \overline{\mathbf{x}}_j}{d_j},
    \label{eq:scaling}
\end{equation}

\noindent where $\overline{\mathbf{x}}_j$ is the mean or expected value of $\mathbf{x}_j$ and $d_j$ is the scaling parameter. Many scaling strategies have been proposed in the literature \cite{Parente2013, VandenBerg2006}, and three popular approaches will be discussed in this work:

\begin{itemize}
    \item \textbf{No scaling}: $d_j=1$. This results in performing PCA on the covariance matrix. Variables with large variances and numerical values will be given higher importance in the decomposition.
    \item \textbf{Auto scaling}: $d_j = \sigma_j$, where $\sigma_j$ is the standard deviation of $\mathbf{x}_j$. This results in performing PCA on the correlation matrix. All variables are standardized to unit variance and are therefore given equal \textit{a priori} importance in the decomposition.
    \item \textbf{Pareto scaling}: $d_j = \sqrt{\sigma_j}$. This is a middle ground option between no scaling and auto scaling. High importance is given to variables with very large variance, but the correlation structure of the dataset stays partially intact.
\end{itemize}

\begin{figure}[h]
	\centering
	\includegraphics[scale=0.43]{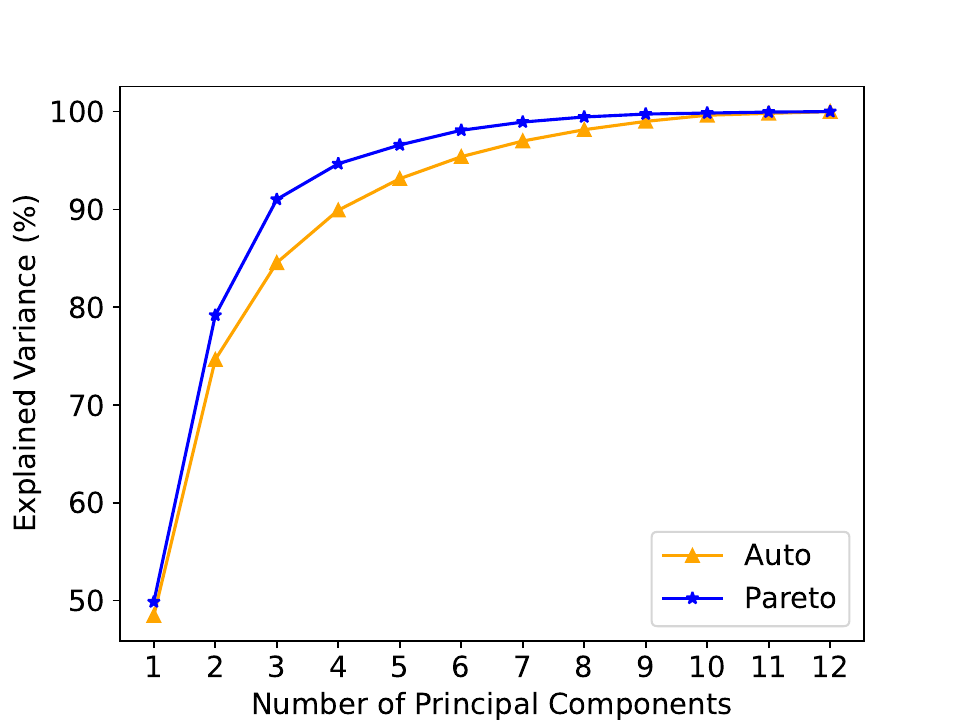}
	\caption{Cumulative variance explained by successive PCs following different data scaling strategies}
	\label{fig:cumulative-explained-variance}
\end{figure}

As discussed above, auto-scaling neglects the fact that the magnitudes of the passive scalars have physical meaning. On the other hand, leaving the data unscaled can excessively bias the low-order PCA model towards a reduced number of variables, leaving important passive scalars unrepresented. Finally, although Pareto scaling also gives higher \textit{a priori} importance to variables with very large variance, the resulting PCA model is likely to be a better representation of the low-rank structure of $\mathbf{X}$ compared to the no-scaling approach because it will be less biased towards a reduced number of passive scalars. \cref{fig:cumulative-explained-variance} shows the cumulative variance explained by successive PCs in a typical dataset consisting of 12 passive scalar time series sampled at a fixed location, following auto-scaling and Pareto scaling of the data. As illustrated, two PCs explain more than 70\% of the total variance in both cases.

\subsection{Biplots}\label{sec:biplots}
A biplot is a graphical representation, typically two-dimensional, of a multivariate dataset following a PCA decomposition \cite{Jolliffe2016, Varmuza2016}. It consists of:
\begin{enumerate}
    \item \textbf{markers} that represent the multivariate observations projected onto the new basis (the PC scores of matrix $\mathbf{Z}_q$ in \cref{eq:pca-truncated-basis-projection}), and
    \item \textbf{vectors} that represent the original variables plotted in PC-space (the columns of matrix $\mathbf{A}_q$ in \cref{eq:pca-truncated-basis-projection}).
\end{enumerate}
Normally, the data is plotted using the first two PCs as coordinates, as they represent the largest proportion of cumulative variance. In the context of this investigation, each marker in the biplot corresponds to one time point sampled from the LES calculation. On the other hand, each vector represents a passive scalar in the two-dimensional PCA model.

As discussed by Jolliffe and Cadima \cite{Jolliffe2016}, the biplot is a useful graphical tool that gives insight into the structure of a multivariate dataset:
\begin{enumerate}
    \item The cosine of the angle between any two vectors (passive scalars) is an approximate measure of their correlation. Two variables pointing in the same direction are strongly positively correlated. If they point in opposite directions, they are strongly anti-correlated. On the other hand, if they are perpendicular, no linear correlation exists between them.
    
    \item The length of a vector is proportional to that variable's contribution to the low-order model. Data scaling (see \cref{sec:data-scaling}) can be employed to increase the relevance of particular variables of interest \textit{a priori}.

    \item Variables (vectors) located in the same area of the plot as a group of observations (markers) tend to have high values for those observations.

    \item Observations (markers) that are close to each other tend to be similar. One can say that proximity in PC-space indicates similarity.
\end{enumerate}

The properties listed above are only exact if all PCs are used. However, because the data is typically plotted in the space spanned by the first two PCs, the more variance that is explained by these components, the more accurate the approximations \cite{Jolliffe2016}.

\begin{figure}[h]
    \centering
    \includegraphics[width=0.6\linewidth]{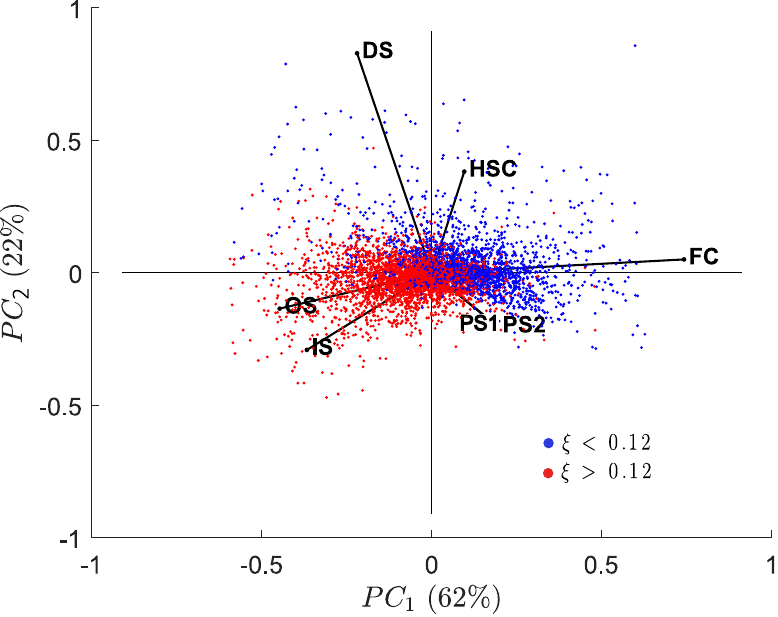}
    \caption{Biplot of a Pareto-scaled matrix of 8 oxidizer passive scalar time series. The temporal observations have been colored by their mixture fraction ($\xi$) value. The percentage of variance explained by each component is specified in brackets.}
    \label{fig:biplot-illustration}
\end{figure}

In order to use a biplot to visually identify the oxidizer streams that contribute the most variance/energy to the local fuel-air mixing process, a strategy that appears reasonable is to perform PCA on the oxidizer passive scalars ($Y_j$) only, omitting the mixture fraction ($\xi$), and to preprocess the oxidizers by performing Pareto scaling (see \cref{sec:data-scaling}). In the biplot, the multivariate observations can then be colored by their instantaneous mixture fraction value, and the length of the vectors can be used as an indicator of the relevance of a particular oxidizer stream to the PCA model (following property (2) above). This is the strategy that has been followed in \cref{fig:biplot-illustration}, which shows a biplot in the space spanned by the first two PCs of the multivariate time series of passive scalars sampled at a spatial location of the experimental test rig. The mixture fraction was employed to split the instantaneous snapshots into two groups according to their numerical value (a value of 0.12, close to the median value, was found to give a good visual split). The first two PCs explain more than 80\% of the total variance. Following property (1) above, the inner swirler (IS) and outer swirler (OS) streams (see \cref{fig:test-rig} for reference) are strongly positively correlated to each other, and are in turn strongly negatively correlated with the film cooling (FC) stream. The fuel-rich cloud of observations (red markers) is clearly aligned with the IS and OS streams, while the FC stream points in the direction of the fuel-lean cloud (blue markers). This indicates that the IS and OS streams have undergone significant mixing with each other and with the fuel stream prior to reaching this spatial location, and are therefore the main fuel carriers to this region of the combustor. The FC stream, on the other hand, has undergone little to no mixing with the fuel stream and directly competes with the IS and OS streams in the local mixing process, contributing towards leaning the mixture. Finally, the dome swirler (DS) stream is perpendicular (i.e. is not linearly correlated) to the IS and OS streams, indicating that it has undergone incomplete mixing with these oxidizer inflows, and by extension with the fuel stream, upstream of the spatial location depicted in the biplot. The Heat Shield Cooling (HSC) flow, and particularly the two Port Streams (PS1 and PS2), have relatively short vectors compared to the other oxidizer streams, indicating that their variance contribution to the mixing process at this particular location is significantly lower than that of the other oxidizer inflows, and therefore play a secondary role in the local fuel-air mixing.

\subsection{K-Means Clustering}\label{sec:k-means-clustering}
\begin{figure}[h]
\centering
\includegraphics[width=0.8\textwidth]{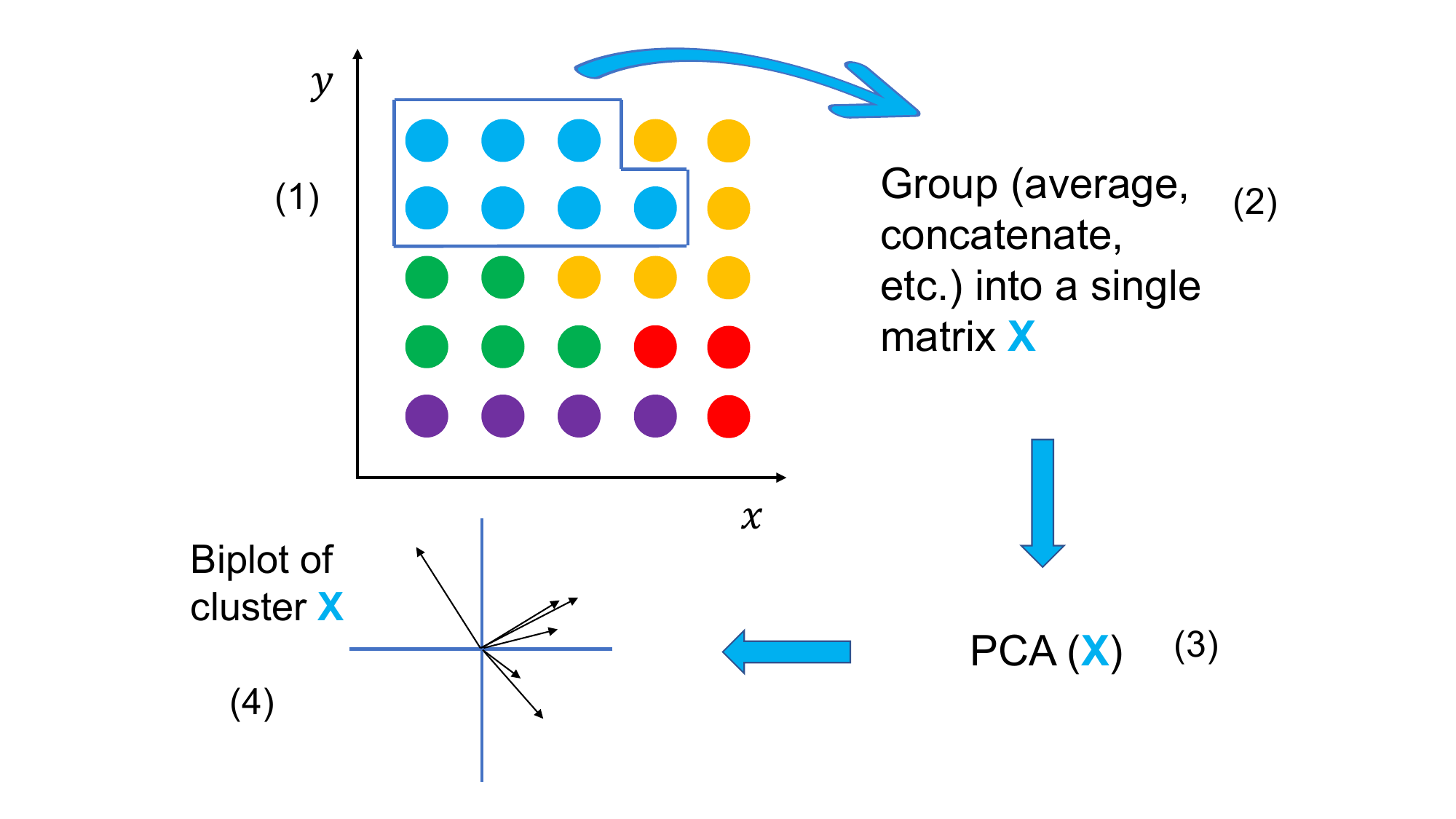}
\caption{Illustration of the clustering of spatial locations, followed by PCA modelling of each cluster and visualization via a biplot. Each point in the $x-y$ grid represents a spatial location where passive scalars have been sampled over time.}
\label{fig:data-sets-clustering-illustration}
\end{figure}

In order to identify spatial locations of the combustor with similar mixing patterns between the fuel and air streams, one can produce multiple spatially-localized PCA models of the time series of fuel and oxidizer stream tracers, and compute their degree of similarity. Locations with similar PCA models can then be clustered together. This approach would allow to identify regions of the combustor with similar mixing trends in an unsupervised fashion, and employ a single biplot to visualize the mixing across a cluster of spatial locations. The concept is illustrated in \cref{fig:data-sets-clustering-illustration}.

Singhal and Seborg \cite{Singhal2005} proposed a modified version of the K-means clustering algorithm which employed a distance metric based on PCA to compute the similarity between multivariate time series in the context of chemometrics. This idea has been borrowed and applied to analyze fuel-air mixing inside gas turbine combustors. The K-means clustering algorithm employs a distance metric to partition a dataset into K clusters, in which objects within the same cluster are as close to each other as possible, and as far from objects in other clusters as possible \cite{Brunton2019}. The distance metric employed by Singhal and Seborg \cite{Singhal2005} is the PCA Similarity Factor between data matrices $\mathbf{X}_1$ and $\mathbf{X}_2$ proposed by Johannesmeyer \cite{Johannesmeyer1999}:

\begin{equation}
	S_{PCA} = \frac{ \sum_{i=1}^{Q} \sum_{j=1}^{Q} \big( l_i^{(1)} l_j^{(2)} \big)  cos^2 \theta_{ij} }
	{ \sum_{i=1}^{Q} l_{i}^{(1)} l_{i}^{(2)} },
	\label{eq:pca-similarity-factor}
\end{equation}

\noindent where $\theta_{ij}$ represents the angle between the $i^{th}$ PC of dataset $\mathbf{X}_1$ and the $j^{th}$ PC of dataset $\mathbf{X}_2$, $l_i^{(1)}$ and $l_j^{(2)}$ are the $i^{th}$ and $j^{th}$ eigenvalues of $\mathbf{X}_1$ and $\mathbf{X}_2$, respectively, and $Q$ is the number of retained principal components in both data matrices. $S_{PCA}$ is conveniently bounded between 0 and 1, where a value of 1 implies that the two PCA models are perfectly similar and a value of 0 implies they are totally dissimilar. Intuitively, $cos^2 \theta_{ij}$ in \cref{eq:pca-similarity-factor} will equal unity if the $i^{th}$ and $j^{th}$ PCs are parallel (i.e. if they are perfectly correlated), and will equal 0 if they are perpendicular (i.e. if no linear correlation exists between them). The PCs are weighted by the variance they explain, thereby reducing the importance of less relevant components in the calculation of $S_{PCA}$. 

\begin{figure}[h]

\begin{subfigure}{.32\textwidth}
        \centering
        \includegraphics[width=1\linewidth]{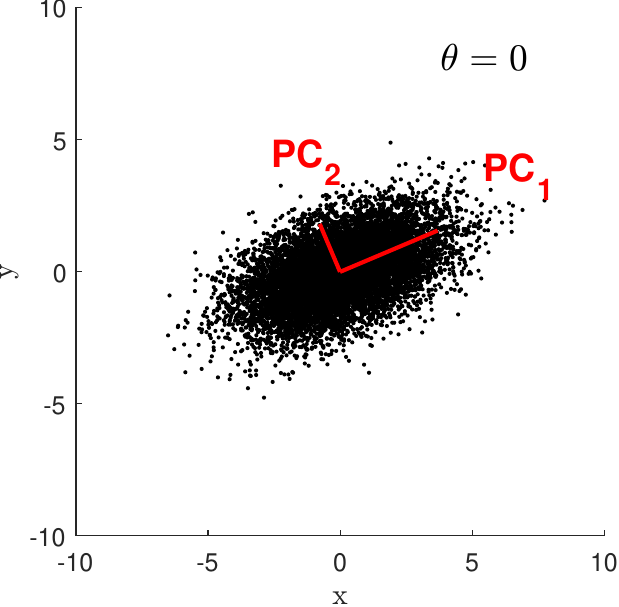}
        \caption{Baseline}
\end{subfigure}
\hfill
\begin{subfigure}{.32\textwidth}
        \centering
        \includegraphics[width=1\linewidth]{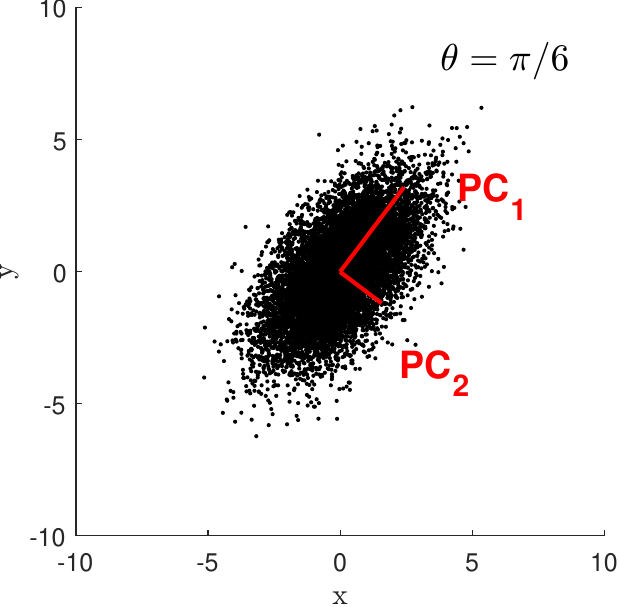}
        \caption{Rotated by $\theta = \pi/6$ \textit{rad}}
\end{subfigure}
\hfill
\begin{subfigure}{.32\textwidth}
        \centering
        \includegraphics[width=1\linewidth]{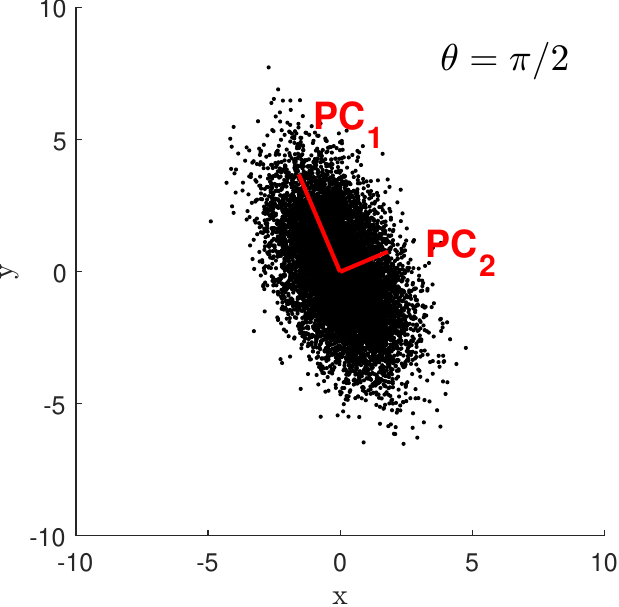}
        \caption{Rotated by $\theta = \pi/2$ \textit{rad}}
\end{subfigure}

\caption{The same bivariate normal distribution rotated by different angles with respect to the baseline, together with its principal components}

\label{fig:pca-similarity-factor-illustration}
\end{figure}

If the dataset consists of a large number of data matrices, a straightforward method to choose $Q$ that allows to easily automate the algorithm is to truncate at a predetermined amount of total variance (e.g. 90\% or 95\%). With this approach, $Q$ should be chosen so that the retained PCs explain at least this predetermined amount of variance in every data matrix (or a large proportion of them, to allow for the possible existence of outlier data matrices that are not well represented by a low-rank structure). This is the strategy that has been followed in this work, given that the number of spatial locations inside the combustion chamber at which passive scalar signals are sampled over time can be in the order of thousands or tens of thousands. The concept behind $S_{PCA}$ is illustrated in \cref{fig:pca-similarity-factor-illustration} for a simple bivariate normal distribution rotated by multiple angles. If only the first PC is retained ($Q=1$), the similarity factors between (a) and (b) or (c) are: $S_{PCA}(a, b) = 0.75$, $S_{PCA}(a, c) = 0$. If both PCs are retained ($Q=2$), the similarity factors are: $S_{PCA}(a, b) = 0.87$, $S_{PCA}(a, c) = 0.48$.

\begin{figure}[h]
\centering
\includegraphics[trim = {3cm, 3cm, 3cm, 3cm}, clip, width=0.8\textwidth]{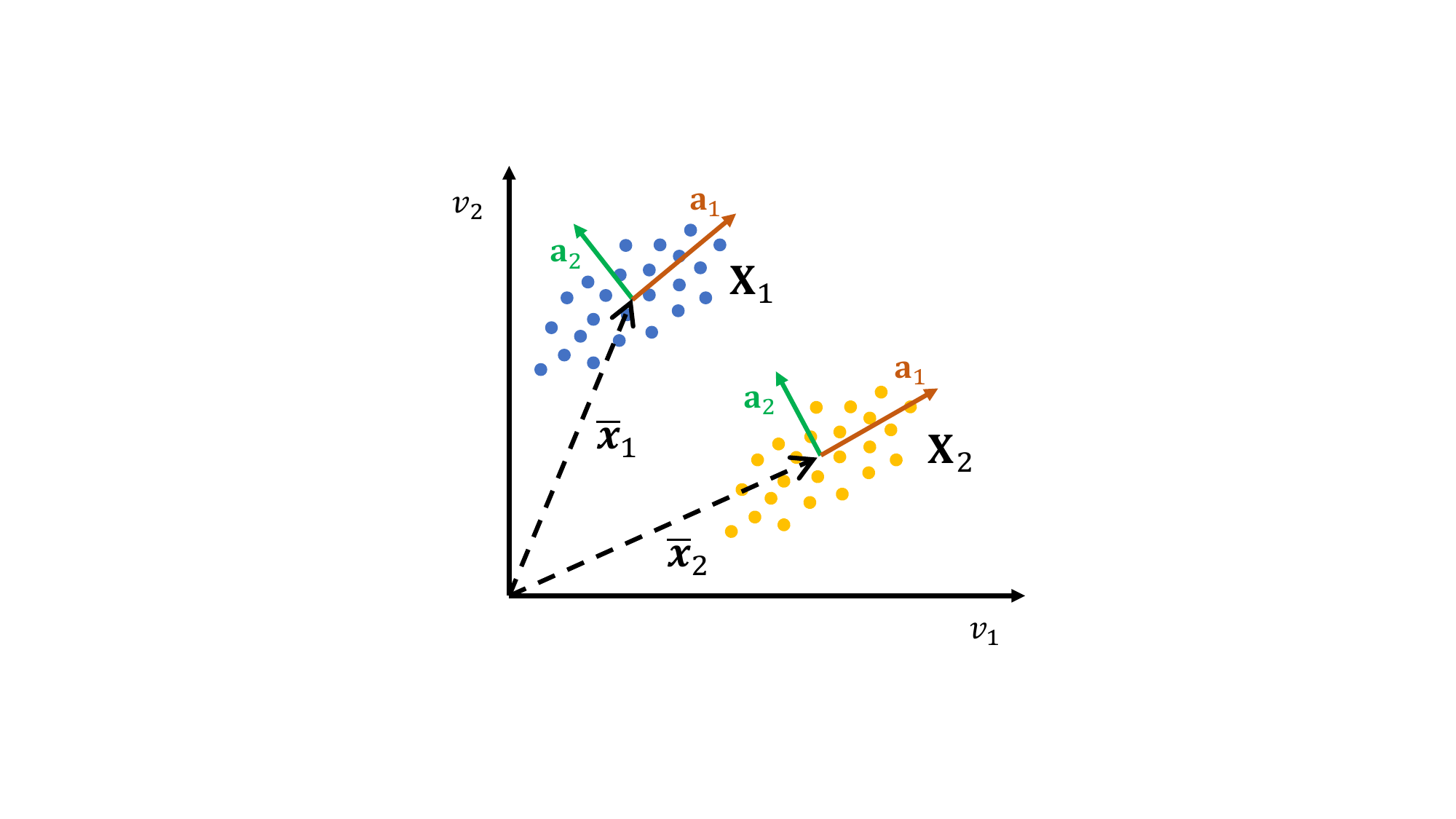}
\caption{Illustration of two bivariate distributions, $\mathbf{X}_1$ and $\mathbf{X}_2$, having a similar spatial orientation (similar PCs $\mathbf{a}_1$ and $\mathbf{a}_2$) but located far apart, i.e. their multivariate means, represented by vectors $\overline{\mathbf{x}}_1$ and $\overline{\mathbf{x}}_2$, are significantly different}
\label{fig:dist-sim-factor-illustration}
\end{figure}

While $S_{PCA}$ measures whether data matrices $\mathbf{X}_1$ and $\mathbf{X}_2$ have similar correlation structures, it does not measure whether their mean value is comparable. Inside the combustion chamber, two spatial locations may have similar mixing trends over time between the multiple air and fuel streams, but the time-averaged value of the oxidizer and fuel streams mass fractions could be different. An example of two bivariate distributions having similar correlation structure but different mean point is illustrated \cref{fig:dist-sim-factor-illustration}. To account for the similarity in the multivariate mean, Singhal and Seborg \cite{Singhal2005} propose an additional metric in their modified version of the K-means clustering algorithm known as the Distance Similarity Factor ($S_{dist}$) \cite{Singhal2002}, which is based on the Mahalanobis distance ($D_M$) of a point from a distribution:

\begin{equation}
	D_M = \sqrt{
	\Bigg( \frac{ \overline{\mathbf{x}}_2 - \overline{\mathbf{x}}_1 }{ \mathbf{d}_1 } \Bigg)
	\: \mathbf{S}_{1}^{\star -1} \:
	\Bigg( \frac{ \overline{\mathbf{x}}_2 - \overline{\mathbf{x}}_1 }{ \mathbf{d}_1 } \Bigg)^T
	},
	\label{eq:mahalanobis-distance}
\end{equation}

\noindent where $\overline{\mathbf{x}}_1$ and $\overline{\mathbf{x}}_2$ are the mean vectors of data matrices $\mathbf{X}_1$ and $\mathbf{X}_2$, respectively, and $\mathbf{d}_1$ is the scaling vector of $\mathbf{X}_1$ (see \cref{eq:scaling}). $\mathbf{S}_{1}^{\star -1}$ is the Moore-Penrose pseudo-inverse of the covariance of $\mathbf{X}_1$ (see \cref{eq:covariance-eigenvalue-decomposition}). $\mathbf{S}_{1}^{\star -1}$ can be calculated using the $Q$ singular values explaining a predetermined amount of total variance. In \cref{eq:mahalanobis-distance}, $\mathbf{X}_1$ acts as the reference distribution against which the mean of $\mathbf{X}_2$ is compared.

\begin{figure}[h]
\centering
\includegraphics[width=0.4\textwidth]{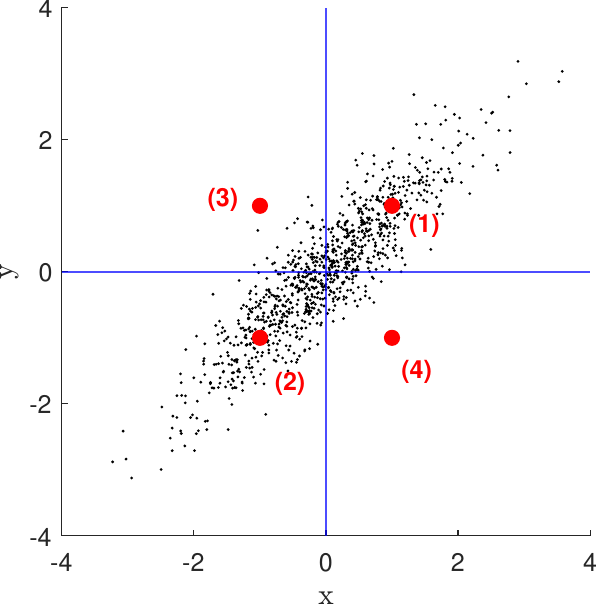}
\caption{Comparison of Mahalanobis and Euclidean distances in a bivariate distribution. The four points have the same Euclidean distance from the mean of the distribution at $(0, 0)$, but points (3) and (4) have a much higher Mahalanobis distance.}
\label{fig:mahalanobis-distance-illustration}
\end{figure}

Employing the Mahalanobis distance takes into account the shape of the data distribution, which the Euclidean distance does not consider. For instance, imagine the distribution shown in \cref{fig:mahalanobis-distance-illustration} represents a bivariate time series sampled at a reference spatial location of a simple combustion system consisting of only two streams (a single fuel inlet and a single oxidizer inlet). Points (1) and (3) represent the mean values at two different but nearby spatial locations. Although both points are the same Euclidean distance away from the mean of the reference distribution, point (1) operates, in an average sense, much closer to the distribution because it is a value that is typically observed at the reference location, whereas point (3) is not. While the example employed here is two-dimensional, this concept can be extended to any number of dimensions (any number of fuel and oxidizer streams). 

Because $D_M$ can be arbitrarily large and it is convenient to have a similarity factor that is bounded between zero and unity, Singhal and Seborg \cite{Singhal2002} suggest performing a monotonic mapping from $D_M$ to $S_{dist}$ using the normal cumulative distribution function with zero mean:

\begin{equation}
	S_{dist} = 2 \,
	\Big[ 1 -
	\frac{1}{\sigma \sqrt{2\pi}} \int_{-\infty}^{D_M} exp \Big( - \frac{z^2}{2 \sigma^2} \Big) dz
	\Big ],
	\label{eq:distance-similarity-factor}
\end{equation}

\noindent where $S_{dist}$ is bounded between zero and unity and a variable standard deviation $\sigma$ allows to control the rate of decay of $S_{dist}$ with increasing $D_M$, which is calculated from \cref{eq:mahalanobis-distance}.

Based on the above discussion, two metrics can be employed to compare the mixing conditions of two spatial locations in a gas turbine combustor. The PCA similarity factor (\cref{eq:pca-similarity-factor}) compares the correlation structure between the fuel and oxidizer streams at the two spatial locations. On the other hand, the distance similarity factor (\cref{eq:distance-similarity-factor}) determines whether the mean operating conditions, in a time-average sense, of the two locations are similar.
In the clustering algorithm the two metrics can be combined into a single similarity factor $S$ through a weighted sum, as proposed by Singhal and Seborg \cite{Singhal2005}:

\begin{equation}
	S = \alpha S_{PCA} + (1 - \alpha) S_{dist},
	\label{equation:combined-similarity-factor}
\end{equation}

\noindent where $\alpha$, bounded between zero and unity, is the weight of the PCA similarity factor in the calculation of the combined metric. Through an appropriate choice of $\alpha$, higher relevance can be given to one similarity factor or the other. It is up to the user to choose the right value of $\alpha$ depending on their application. Singhal and Seborg \cite{Singhal2005} report good results with $\alpha=0.67$, although this is highly dependent on the dataset the algorithm is applied to. In the application proposed in the present work, the main objective is to find spatial locations where the correlation between streams is similar, so that a single biplot can be used to visualize the mixing across a cluster of spatial locations. Therefore, values of $\alpha>0.5$ seem reasonable.

\subsection{Data Sampling}\label{sec:data-sampling}

\begin{figure}[h]
\begin{subfigure}{.5\textwidth}
        \centering
        \includegraphics[width=0.9\linewidth]{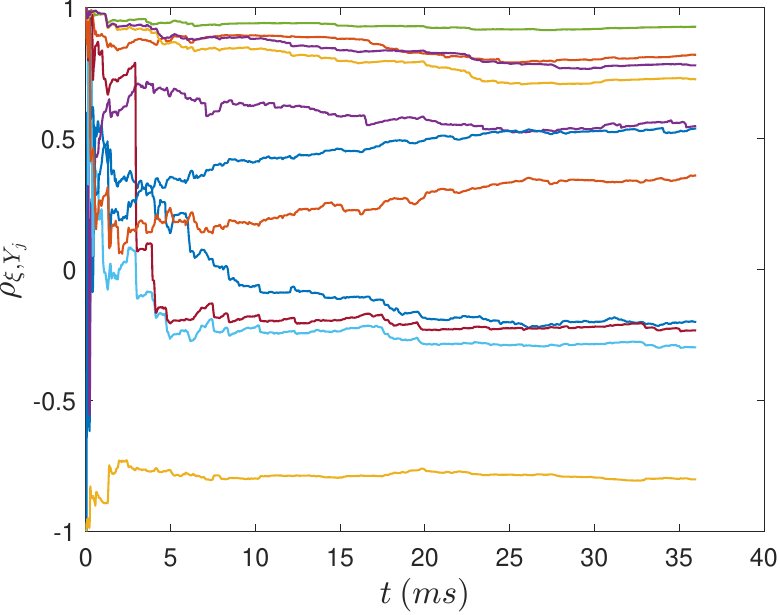}
        \caption{Location 1 ($I = 48\%$)}
\end{subfigure}
\begin{subfigure}{.5\textwidth}
        \centering
        \includegraphics[width=0.9\linewidth]{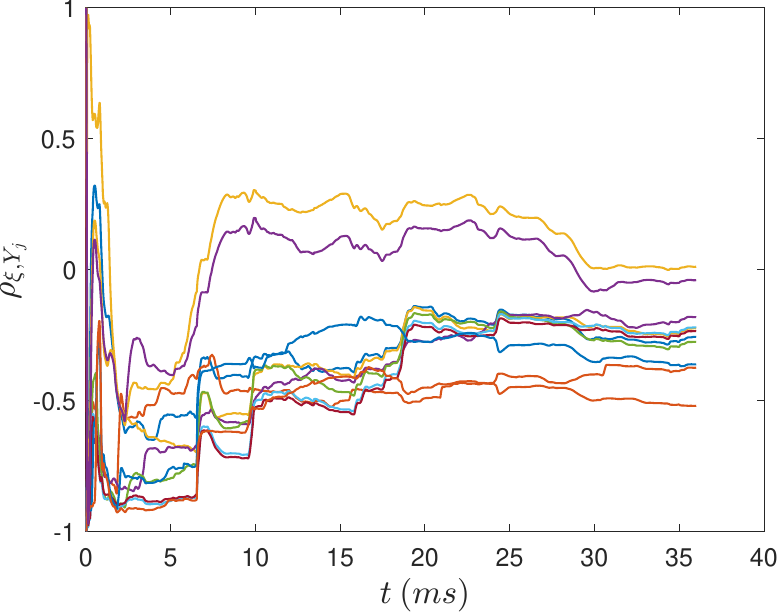}
        \caption{Location 2 ($I = 126\%$)}
\end{subfigure}

\caption{Statistical convergence of the correlation coefficients of the mixture fraction with the oxidizer passive scalars at two spatial locations of the experimental test rig with different turbulence intensity $I$.}

\label{fig:statistical-convergence-of-correlations}
\end{figure}

In order to apply the statistical methodology described above, time series of the oxidizer passive scalars and the mixture fraction must be sampled across $M$ spatial locations in regions of interest of the combustor during the LES calculation. The time series sampled at each spatial location are arranged into a tall matrix $\mathbf{X}$ of size $(N \times P)$, where $N$ is the number of snapshots or temporal observations and $P$ is the number of passive scalars. In order to calculate statistically meaningful correlations between passive scalars it is necessary to sample a sufficiently large number of statistically independent snapshots. As a general rule, the higher the statistical moment, the greater the fluctuation level of a random variable relative to its mean, and the greater its deviation from Gaussian behavior, the larger $N$ needs to be to achieve a sufficiently high statistical accuracy \cite{Pope2000}. \cref{fig:statistical-convergence-of-correlations} shows the convergence of the correlation coefficients of the mixture fraction with the oxidizer passive scalars at two locations of the experimental test rig (see \cref{sec:investigated-geometries}) as a function of sampling time. The two spatial locations have significantly different turbulence intensity, which is given by $I = u_{rms}/\overline{u}$, where $u_{rms}$ and $\overline{u}$ are the root-mean-square and mean velocities, respectively. At location 1, convergence is reached relatively quickly. However, at location 2, which is located close to the central recirculation zone, it takes approximately 30 $ms$ to reach convergence, which is significant considering that the flow-through time (the average time it takes for a fluid particle to cross the combustor from inlet to outlet) is approximately 10 $ms$ for this test case.

\section{Results}\label{sec:results}

\subsection{Choosing the number of clusters}\label{sec:choosing-number-clusters}
One of the first questions to be answered when applying the K-means clustering algorithm to a particular dataset is the number of clusters ($K$) to choose.
In this work, following the discussion in \cref{sec:k-means-clustering}, a cluster may be seen as a ``mixing region", where the combustor spatial locations belonging to the same cluster have similar mixing characteristics (as described by the relationship between the fuel and oxidizer stream tracers, which are sampled over time and arranged into a tall matrix $\mathbf{X}$).

To choose an appropriate $K$, Singhal and Seborg \cite{Singhal2005} define a cost function based on the dissimilarity factor:

\begin{equation}
	d_{i,m} = 1 - S_{i,m},
    \label{eq:dissimilarity-factor}
\end{equation}

\noindent where $S_{i,m}$ is the similarity factor (\cref{equation:combined-similarity-factor}) between the $m^{th}$ data instance (in our case, the $\mathbf{X}_m$ matrix at the $m^{th}$ spatial location) and the $i^{th}$ cluster $\mathbf{C}_i$ it belongs to, and $d_{i,m}$ is the corresponding dissimilarity factor. The cost function is then defined as an average of $d_{i,m}$ across all clusters:

\begin{equation}
	J(K) = \frac{1}{M} \sum_{i=1}^{K}\sum_{\mathbf{X}_m \in \mathbf{C}_i} d_{i,m}
	\label{equation:global-dissimilarity},	
\end{equation}

\noindent where $M$ is the total number of data instances (spatial locations) and $K$ is the number of clusters. The clustering algorithm can then be run with an increasing number of clusters and the change of $J(K)$ with $K$ can be monitored. A similar approach was followed by D'Alessio et al. \cite{DAlessio2020} in their application of Local PCA for the thermochemical analysis of turbulent reacting jets: plotting the reconstruction error (their particular cost function) as a function of the number of clusters and identifying the location where the curve starts to asymptote.\par

\begin{figure}[h]
	\centering
	\includegraphics[scale=0.45]{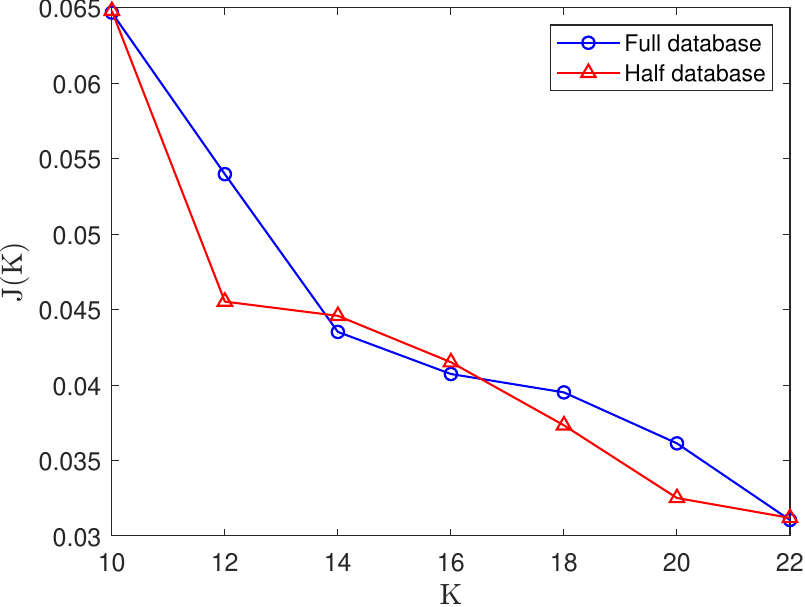}
	\caption{Cost function $J(K)$ as a function of the number of clusters $K$ using (i) all spatial locations and (ii) half of the spatial locations (randomly sampled) across the mid-plane of the experimental test rig}
	\label{fig:find-number-clusters}
\end{figure}

\cref{fig:find-number-clusters} shows $J(K)$ as a function of the number of clusters $K$ for data sampled from the experimental test rig simulation (see \cref{sec:investigated-geometries}).
As expected, $J(K)$ decreases as $K$ is increased because the clusters become smaller and internally more homogeneous. We would also expect the curve to eventually asymptote if $K$ was increased further. In the context of this work, the main objective of the clustering approach is to reduce the large spatial dimensionality of the mixing problem so that a single biplot (\cref{sec:biplots}) can be employed to visually examine the fuel-air mixing patterns across a region of the combustor. Increasing $K$ will provide a more optimized decomposition of the dataset, but will also increase the number of biplots that need to be analyzed.

\subsection{Robustness to initialization}\label{sec:robustness-initialisation}

The K-means clustering algorithm can converge to a solution that is a local (non-global) minimum \cite{Brunton2019}.
This means that the final solution may be sensitive to the user-provided initialization.
Ideally, the converged solution should be reasonably independent from the initial guess provided by the user, and the algorithm should identify the same mixing regions in the combustor regardless of the initialization.

\begin{figure}[h]
\begin{subfigure}{.5\textwidth}
	\centering
	\includegraphics[trim={1cm, 1cm, 1cm, 1cm}, clip, width=1.1\linewidth]{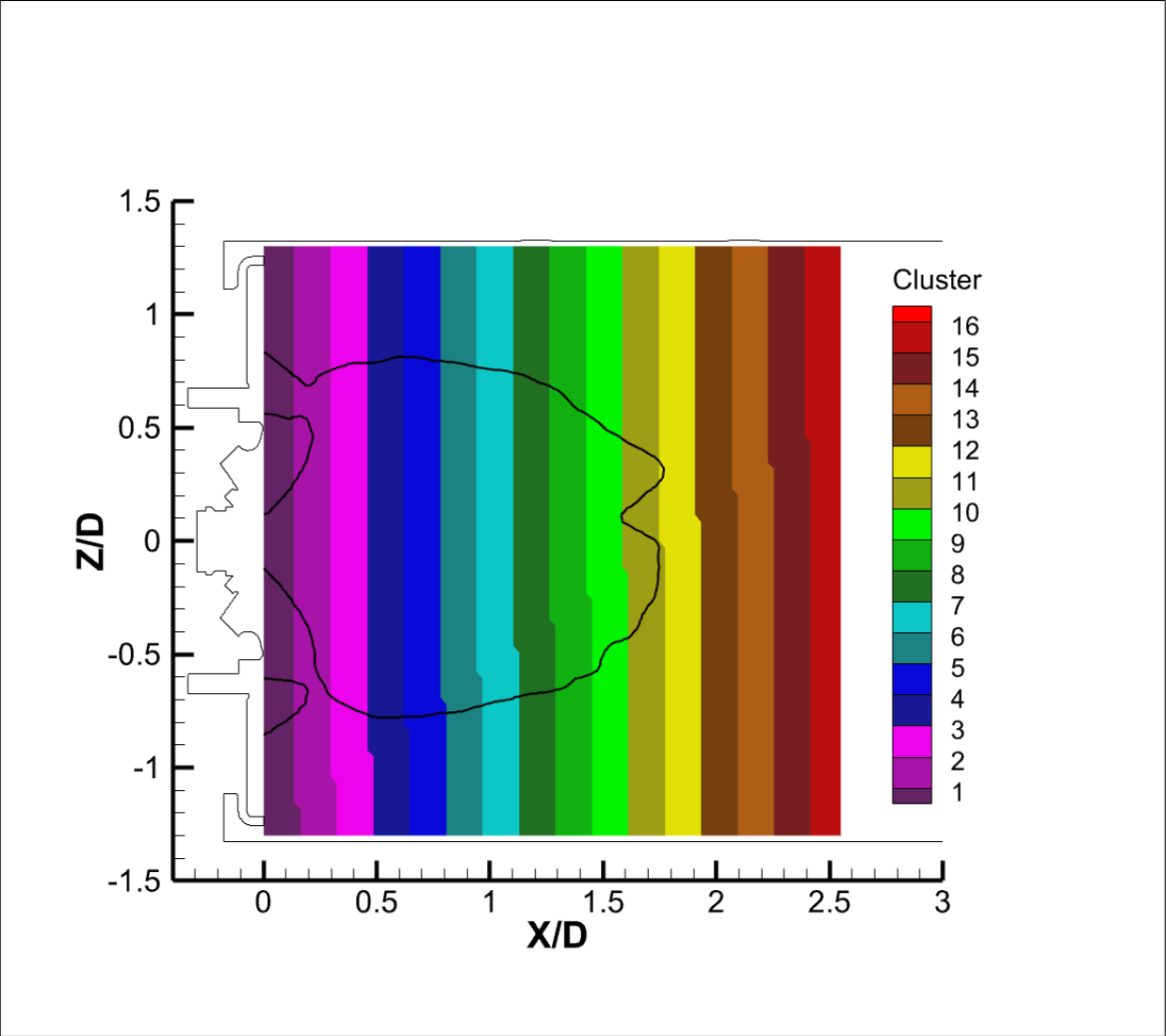}
	\caption{Initialization 1}
    \label{subfig:initialisation-1}
\end{subfigure}
\begin{subfigure}{.5\textwidth}
	\centering
	\includegraphics[trim={1cm, 1cm, 1cm, 1cm}, clip, width=1.1\linewidth]{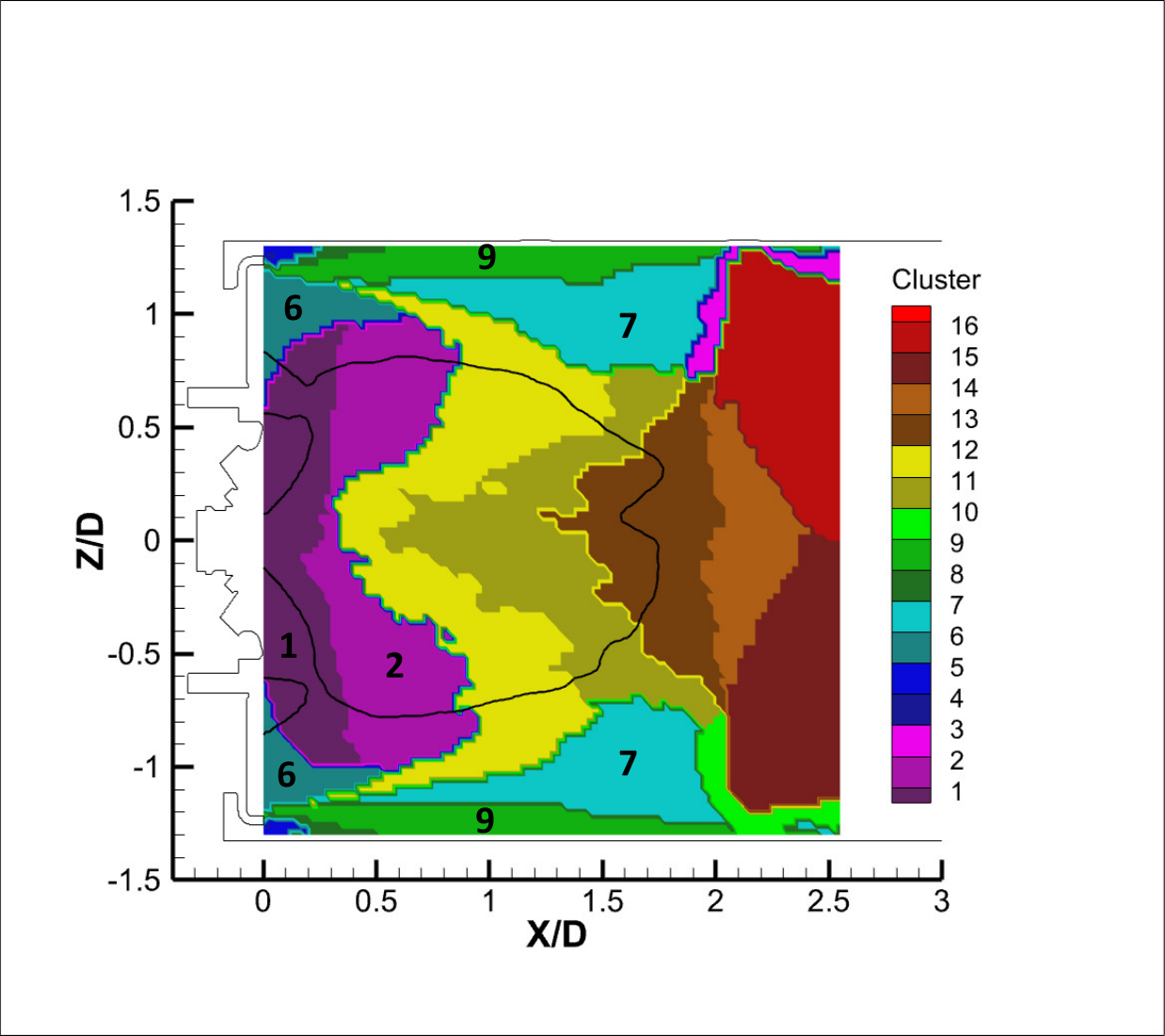}
	\caption{Converged solution of initialization 1}
	\label{subfig:converged-solution-initialisation-1}
\end{subfigure}
\newline
\begin{subfigure}{.5\textwidth}
	\centering
	\includegraphics[trim={1cm, 1cm, 1cm, 1cm}, clip, width=1.1\linewidth]{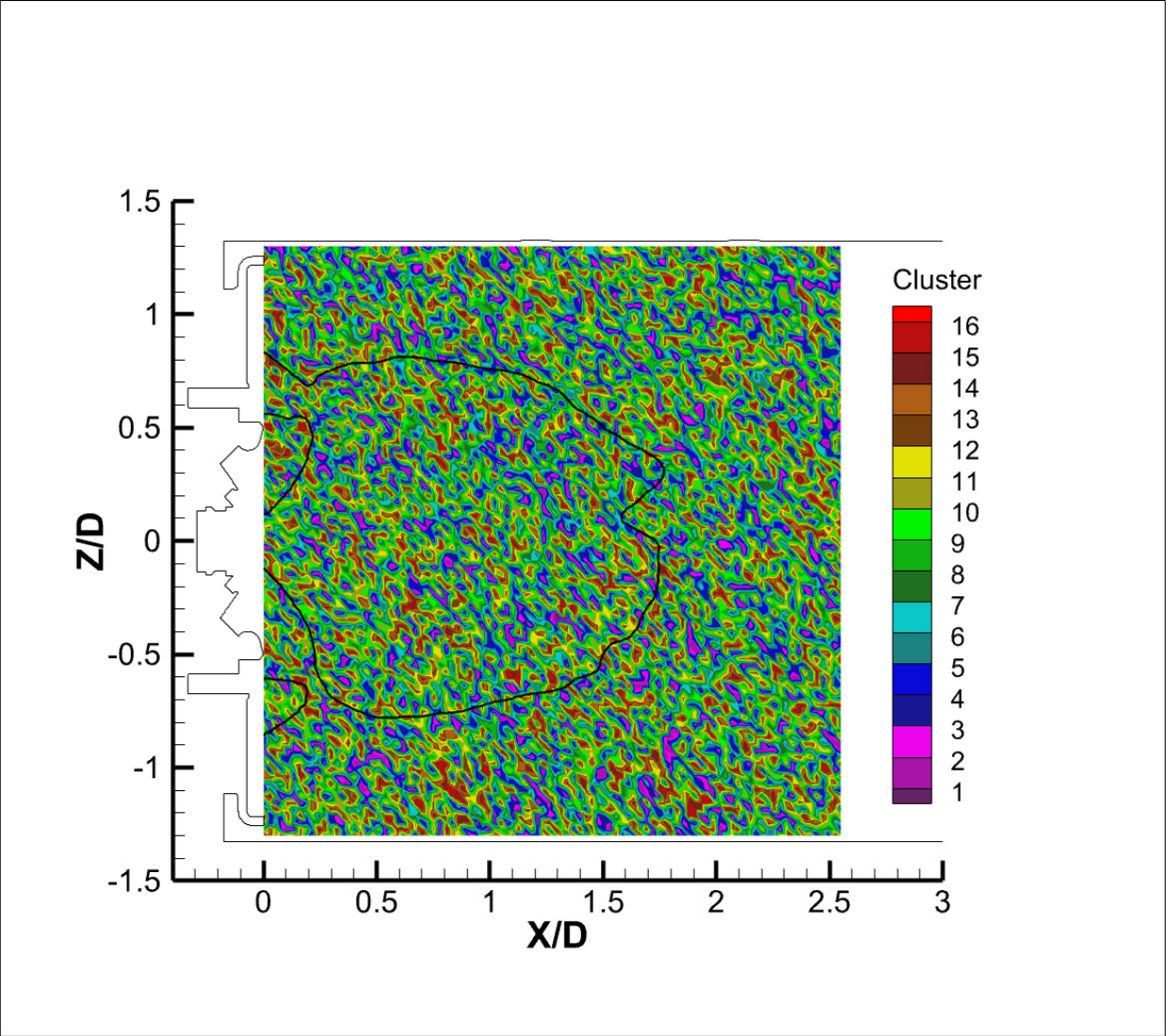}
	\caption{Initialization 2}
\end{subfigure}
\begin{subfigure}{.5\textwidth}
	\centering
	\includegraphics[trim={1cm, 1cm, 1cm, 1cm}, clip, width=1.1\linewidth]{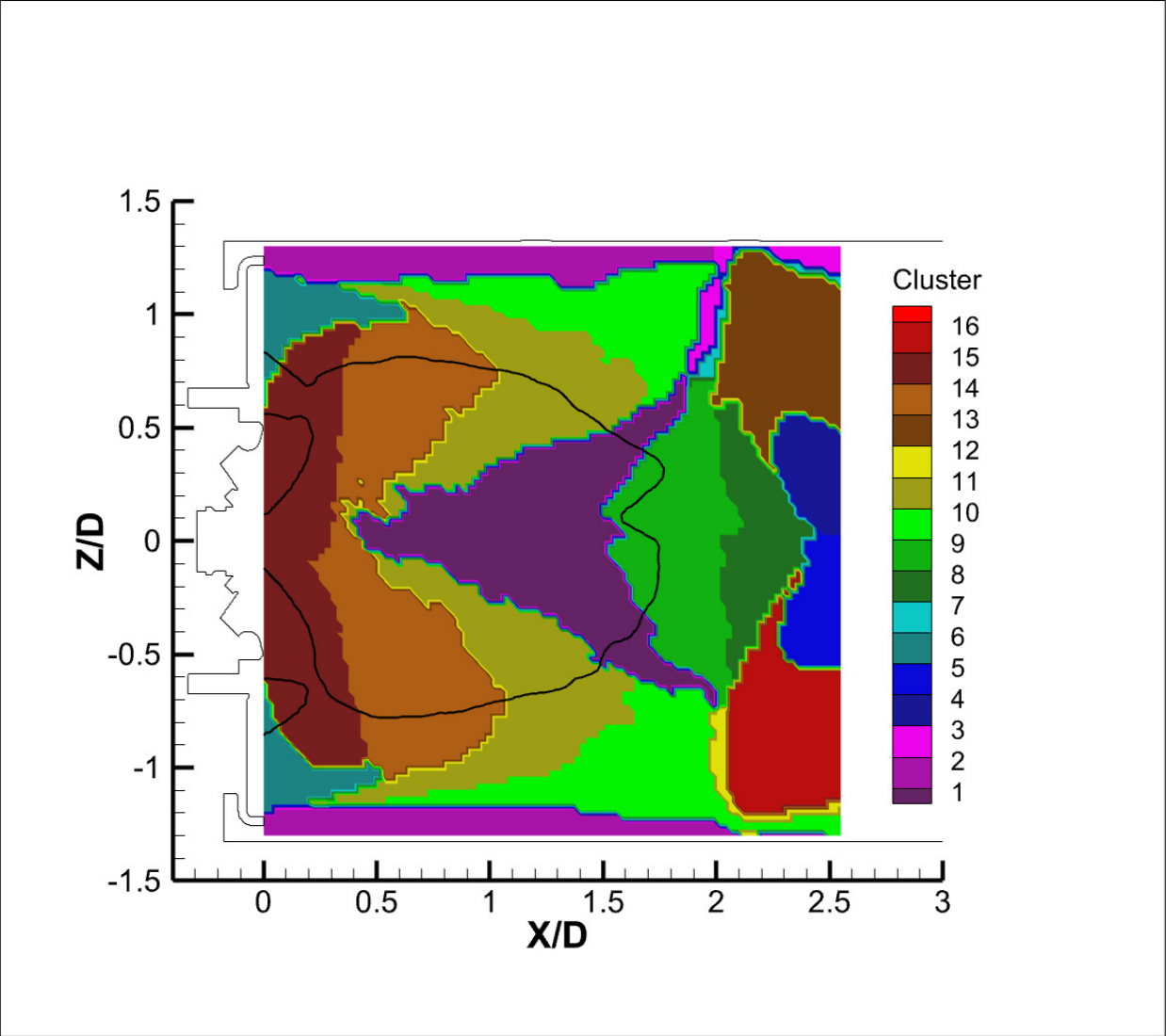}
	\caption{Converged solution of initialization 2}
\end{subfigure}

\caption{Converged solutions across the mid-plane of the experimental rig for two different algorithm initializations with $K = 16$ and $S_{PCA}$ weighting $\alpha=1$.
Every color represents a different cluster.
The black isoline is the zero mean axial velocity level ($\overline{u}_x = 0$). The coordinates have been normalized by one injector diameter $D$.}

\label{fig:clustering-robustness-to-initialisation}
\end{figure}

\cref{fig:clustering-robustness-to-initialisation} shows converged solutions of the clustering algorithm for the mid-plane of the experimental test rig (see \cref{fig:test-rig} for reference) for two different algorithm initializations. The data, consisting of fuel and oxidizer tracers sampled over time at each spatial location, was mean-centered and Pareto-scaled in a pre-processing step (see \cref{sec:data-scaling}), thus giving more relevance in the low-order PCA models to the streams with high variance. In this case the algorithm was run with $\alpha=1$ (see \cref{equation:combined-similarity-factor}), i.e. only similarity in the correlation structure, and not in the mean operating point, was considered. The zero mean axial velocity level is also shown in \cref{fig:clustering-robustness-to-initialisation} to help identify regions of flow recirculation. As illustrated, despite the fact that the two initializations are completely dissimilar, the converged solutions are almost the same. Additionally, the converged solutions agree well with the flow physics inside a gas turbine combustor. For instance, clusters 1 and 2 in \cref{subfig:converged-solution-initialisation-1} are dominated by the injector flow and the transition from cluster 1 to cluster 2 is dominated by the completion of spray evaporation as well as the completion of the mixing between the inner and outer swirler streams.
Cluster 9 is clearly dominated by the wall film cooling flow. The reader is reminded that K-means clustering is a type of unsupervised machine learning, and that the algorithm has no information about the spatial distribution of the data instances.

\begin{figure}[h]
\begin{subfigure}{.5\textwidth}
	\centering
	\includegraphics[width=1\linewidth]{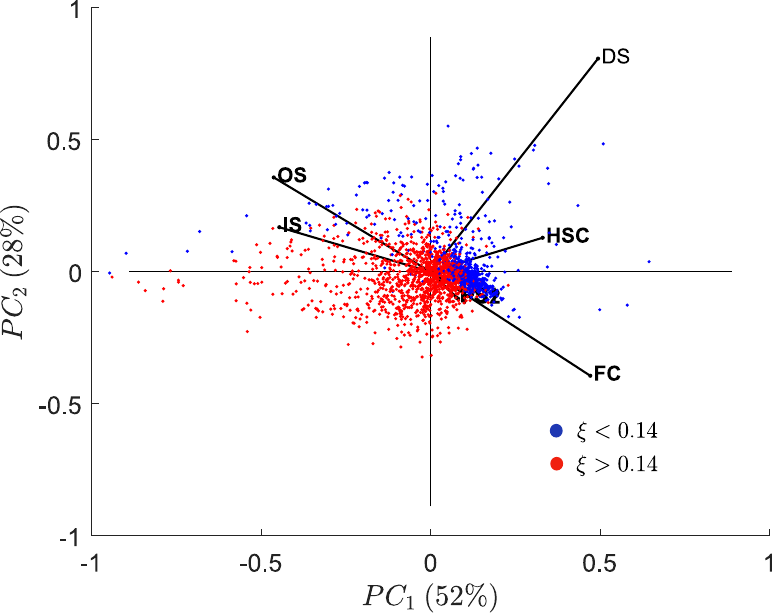}
	\caption{Cluster 2}
	\label{subfig:biplot-cluster-2}
\end{subfigure}
\begin{subfigure}{.5\textwidth}
	\centering
	\includegraphics[width=1\linewidth]{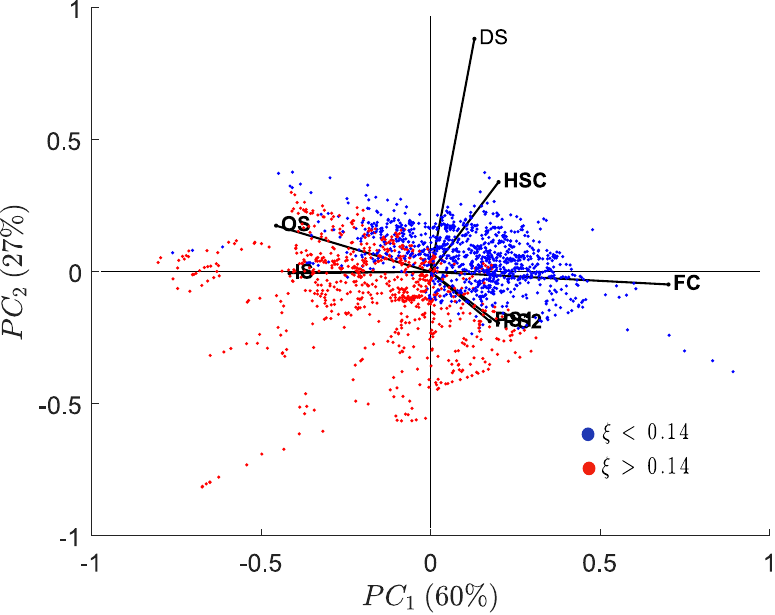}
	\caption{Median point of cluster 2}
\end{subfigure}
\vspace{1cm}
\newline
\begin{subfigure}{.5\textwidth}
	\centering
	\includegraphics[width=1\linewidth]{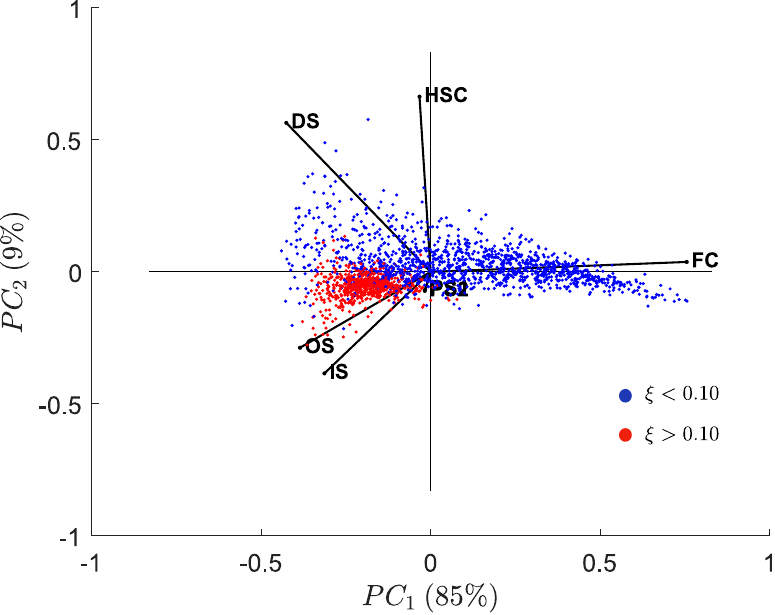}
	\caption{Cluster 7}
	\label{subfig:biplot-cluster-7}
\end{subfigure}
\begin{subfigure}{.5\textwidth}
	\centering
	\includegraphics[width=1\linewidth]{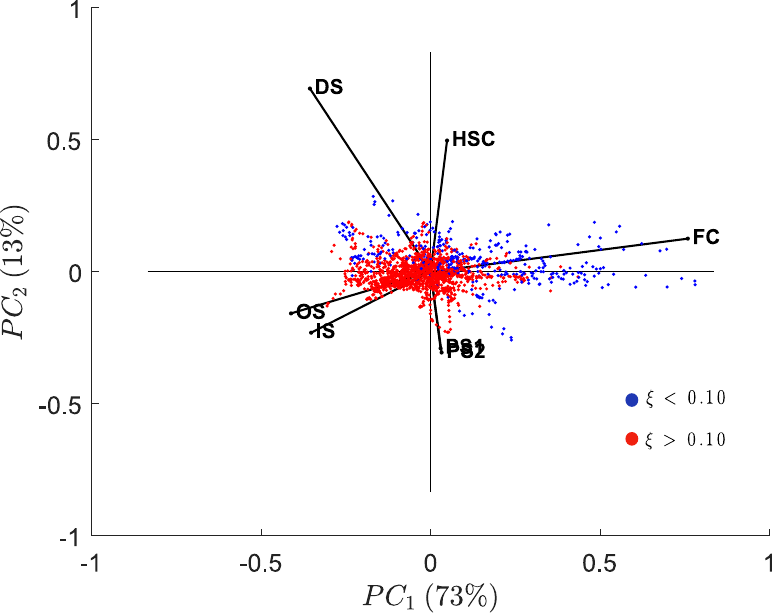}
	\caption{Median point of cluster 7}
\end{subfigure}

\caption{Biplots of clusters 2 \& 7 and of their corresponding median point.
The cluster numbering follows \cref{subfig:converged-solution-initialisation-1}.
The variance explained by each principal component is indicated in brackets.
The data was Pareto-scaled prior to PCA.}

\label{fig:biplots-clusters-2-and-7}
\end{figure}

\cref{fig:biplots-clusters-2-and-7} illustrates the biplots of clusters 2 and 7 shown in \cref{subfig:converged-solution-initialisation-1}, as well as the biplots of the median spatial location of each cluster. The biplots are plotted in the space spanned by the first two principal components (which represent at least 80\% of the total variance in all cases), following the strategy described in \cref{sec:biplots}. Although some small differences exist, the biplots of the clusters and of their respective median points are alike, indicating the similarity in their covariance structure. This proves the usefulness of the proposed implementation of the K-means clustering algorithm: a cluster consists of spatial locations with similar mixing characteristics (covariance structure between stream tracers), so that a single biplot is sufficient to analyze the mixing across all the spatial locations that belong to it. \cref{subfig:biplot-cluster-2,subfig:biplot-cluster-7} show that, despite the physical proximity of the dome swirler (DS) inlet to the inner swirler (IS) and outer swirler (OS) inlets (see \cref{fig:test-rig}), the dome swirler flow struggles to complete its mixing with the inner and outer swirler flows, which are the main fuel carriers (DS is perpendicular to IS and OS, indicating lack of linear correlation). This is the case even at cluster 7, which is primarily located more than 1.5 injector diameters downstream of the injector outlet. \cref{subfig:biplot-cluster-2,subfig:biplot-cluster-7} also indicate that the film cooling (FC) flow plays a relevant role leaning the fuel-air mixture in both clusters, as it has a large magnitude vector which points in the direction of the fuel-lean cloud of observations.

\FloatBarrier

\subsection{Solution convergence}\label{sec:solution-convergence}
\begin{figure}[h]
    \begin{subfigure}{.5\textwidth}
		\includegraphics[width=.95\linewidth]{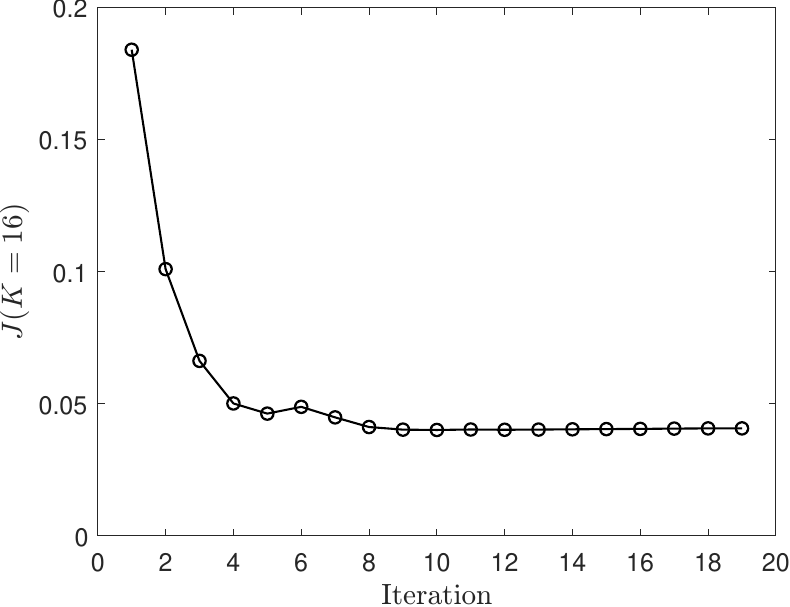}
		\caption{Cost function $J(K)$ vs iteration number}
		\label{subfig:dissimilarity-global}
	\end{subfigure}
	\begin{subfigure}{.5\textwidth}
		\includegraphics[width=.95\linewidth]{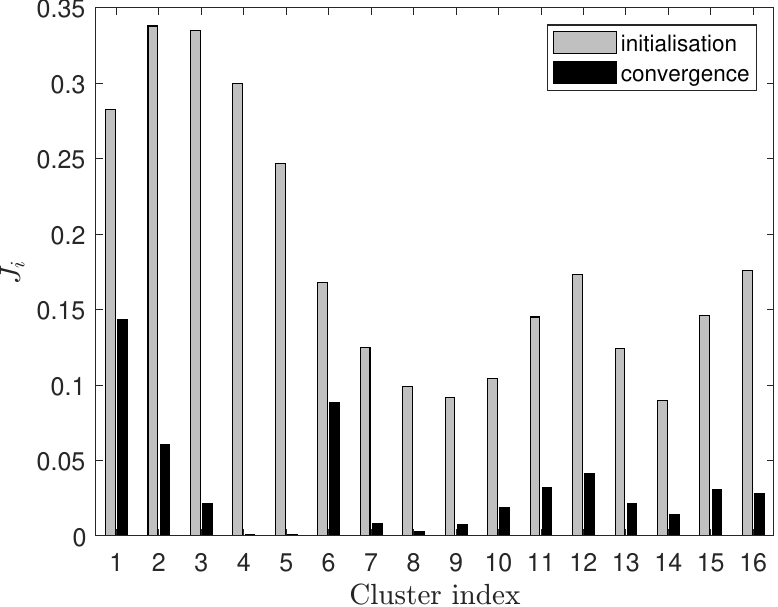}
		\caption{Average within-cluster dissimilarity}
		\label{subfig:dissimilarity-within-cluster}
	\end{subfigure}
\caption{The cost function $J(K)$ as a function of the number of algorithm iterations and the within-cluster dissimilarity $J_i$ at initialisation and at convergence}	
\end{figure}

One approach to monitor the convergence of the K-means clustering algorithm is to plot the cost function $J(K)$ (\cref{equation:global-dissimilarity}) as a function of the number of iterations. This is illustrated in \cref{subfig:dissimilarity-global} for the converged solution shown in \cref{subfig:converged-solution-initialisation-1} with $K=16$ clusters. The algorithm is able to achieve convergence after approximately 10 iterations following the initialization of \cref{subfig:initialisation-1}, on a table-top workstation in less than 5 minutes. This particular dataset consisted of $9000$ sampled spatial locations, with a matrix $\mathbf{X}$ of size ($2000 \times 8$) at each spatial location.

To quantify the homogeneity of a given cluster at convergence, one can calculate the average within-cluster dissimilarity for the $i^{th}$ cluster as:

\begin{equation}
	J_i = \frac{1}{M_i} \sum_{\mathbf{X}_m \in \mathbf{C}_i} d_{i,m},
	\label{eq:within-cluster-dissimilarity}
\end{equation}

\noindent where $d_{i,m}$ is the dissimilarity factor (\cref{eq:dissimilarity-factor}) between the $m^{th}$ data instance $\mathbf{X}_m$ and the $i^{th}$ cluster $\mathbf{C}_i$, and $M_i$ is the number of data instances in the $i^{th}$ cluster. \cref{subfig:dissimilarity-within-cluster} shows the average within-cluster dissimilarity across all 16 clusters at initialization and at convergence. As illustrated, all clusters are considerably inhomogeneous at initialization, as would be expected. $J_i$ drops for all clusters as the algorithm iterates and the clusters are filled with spatial locations with similar correlation structures between stream tracers. At convergence, clusters 1 and 6 are considerably inhomogeneous relative to the other clusters. These correspond to the region just downstream of the injector exit plane and the mixing zone along the chamber back wall (see \cref{subfig:converged-solution-initialisation-1}). The high value of $J_i$ for cluster 1 may be explained by the evaporating spray, the incomplete mixing between the injector swirler flows and the gaseous fuel, and the presence of strong flow recirculation. In the case of cluster 6, in addition to flow recirculation, the mixture inhomogeneity may be explained by the shear-driven mixing layer between the dome swirler flow, the heat shield cooling and the film cooling flows (see \cref{fig:test-rig} for reference). All of these processes lead to strong spatial gradients in the mixture fraction and oxidizer passive scalar fields, resulting in a comparatively inhomogeneous cluster relative to the other clusters.

\subsection{Similarity in mean operating point}\label{sec:similarity-mean-point}

\begin{figure}
	\begin{subfigure}{.5\textwidth}
		\includegraphics[trim={1cm, 1cm, 1cm, 1cm}, clip, width=1.1\linewidth]{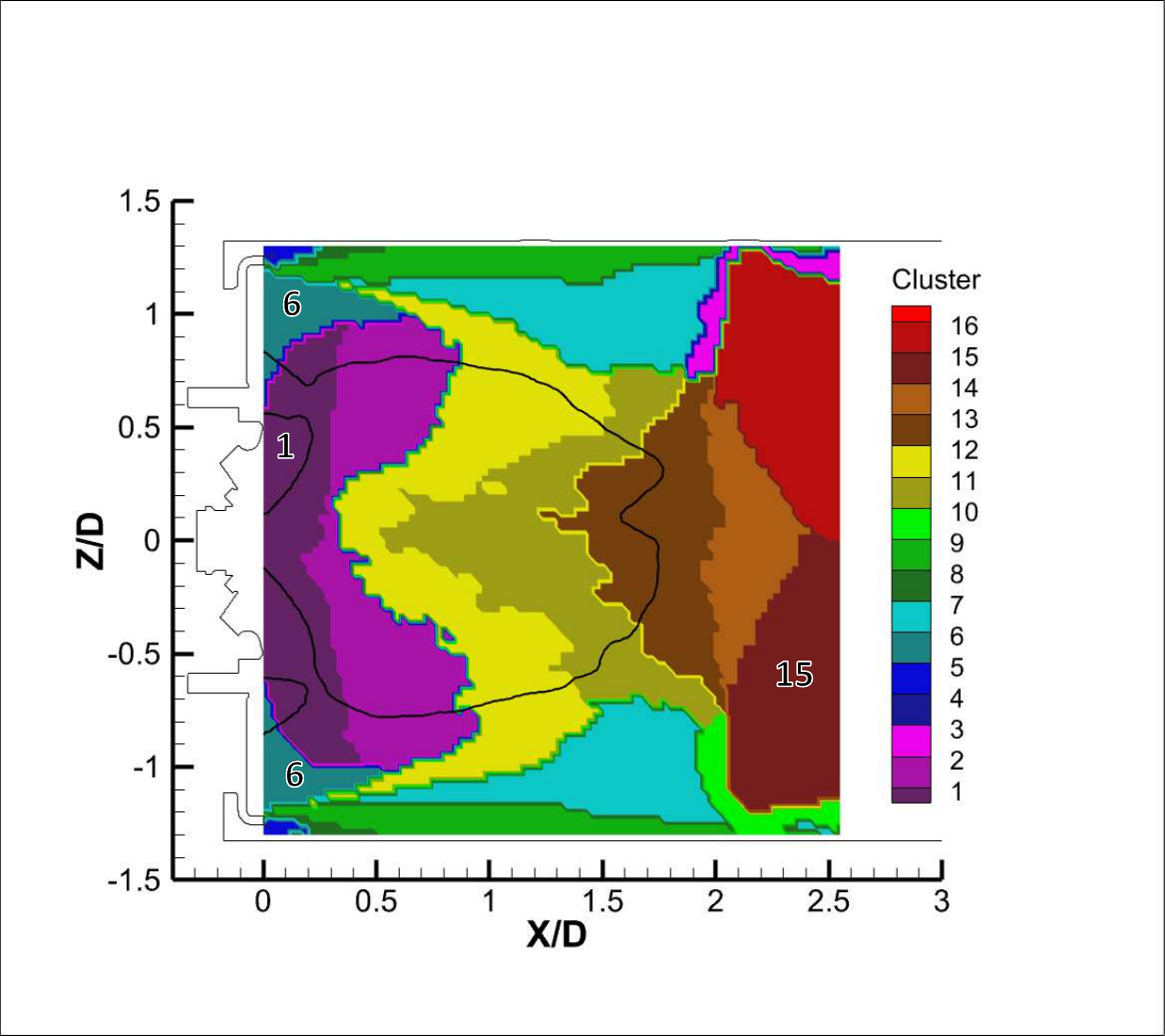}
		\caption{$\alpha = 1$}
		\label{subfig:clusters-alpha-1}
	\end{subfigure}
	\begin{subfigure}{.5\textwidth}
		\includegraphics[trim={1cm, 1cm, 1cm, 1cm}, clip, width=1.1\linewidth]{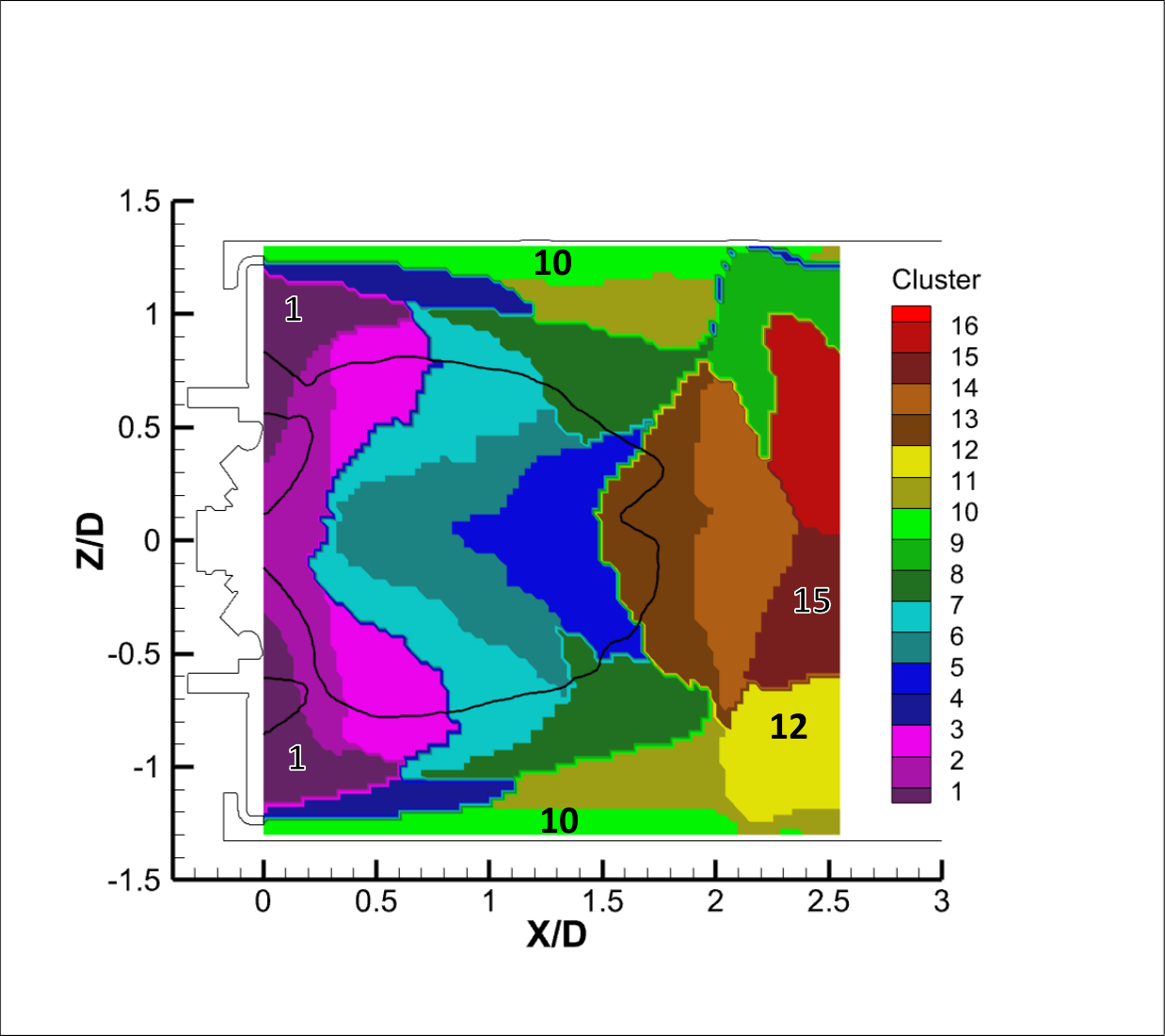}
		\caption{$\alpha = 0.67$}
		\label{subfig:clusters-alpha-0.67}
	\end{subfigure}
\caption{Converged solutions of the experimental rig data with $K=16$ clusters with two different $S_{PCA}$ weighting factors ($\alpha$). The black isoline is the zero mean axial velocity level ($\overline{u}_x = 0$). The coordinates have been normalized by one injector diameter $D$.}
\label{fig:clustering-effect-of-alpha}
\end{figure}

The results presented in \cref{fig:clustering-robustness-to-initialisation} correspond to a $S_{PCA}$ weighting factor of $\alpha=1$ (\cref{equation:combined-similarity-factor}). In other words, the similarity in the mean operating point between spatial locations was not taken into account in the K-means clustering algorithm. \cref{fig:clustering-effect-of-alpha} shows the converged solutions of the experimental test rig mid-plane with $\alpha=1$ and $\alpha=0.67$. Both iterations of the algorithm were started from the same initialization and run until converged. A value of $\alpha = 0.67$ gives the PCA similarity factor a weight of $2/3$ in the clustering algorithm cost function, while the distance similarity factor, which measures the similarity in the multivariate mean of the oxidizer passive scalars and mixture fraction time series, is given a weight of $1/3$.

As illustrated in \cref{fig:clustering-effect-of-alpha}, although the identified patterns are similar, incorporating $S_{dist}$ in the K-means clustering algorithm has an important influence on the solution. For instance, cluster 1 in \cref{subfig:clusters-alpha-0.67} extends radially-inward into the fuel injector exit plane, compared to cluster 6 in \cref{subfig:clusters-alpha-1}. This may be explained by the dominance of the dome swirler flow mass fraction across this cluster (see \cref{fig:test-rig} for reference). Due to the steep metal angle of the injector assembly at the exit plane, the dome swirler flow undergoes a sudden outward radial expansion once it exits the injector, and tends to remain attached to the combustor back wall, increasing its mass fraction across cluster 1 in \cref{subfig:clusters-alpha-0.67}. While the correlation structure between stream tracers is likely to be more homogeneous within clusters 1 and 6 in \cref{subfig:clusters-alpha-1}, $S_{PCA}$ alone ($\alpha$ = 1) cannot capture the dominance of the dome swirler flow across cluster 1 of \cref{subfig:clusters-alpha-0.67} in terms of its mean mass fraction. A similar situation is encountered across cluster 10 in \cref{subfig:clusters-alpha-0.67}, whose composition is dominated by the film cooling flow mass fraction, which is designed to stick to combustor wall to prevent damage from overheating. With $\alpha=1$, this region of the combustor is split into multiple clusters in the vicinity of the film cooling inlet due to the somewhat different correlation structures that result from the shear-driven mixing with the heat shield cooling and dome swirler flows. However, because the mean mass fraction is nevertheless dominated by the film cooling flow, the clustering algorithm identifies this region as a unique cluster when $\alpha=0.67$. Another interesting difference between $\alpha=1$ and $\alpha=0.67$ happens at cluster 15 of \cref{subfig:clusters-alpha-1}, which is split into clusters 12 and 15 in \cref{subfig:clusters-alpha-0.67}. Analysis of the time-averaged passive scalars across clusters 12 and 15 of \cref{subfig:clusters-alpha-0.67} shows that the mass fraction of the dilution port flow emanating from the top wall (at $x/D \approx 2.1$, $z/D = +1.3$) is much higher in cluster 15 than in cluster 12. This may be explained by the proximity of cluster 15 to the impingement point between the top and bottom dilution port jets, close to the chamber centerline at $z/D=0$.

\subsection{Analysis of the detailed single sector combustor }\label{sec:analysis-detailed-combustor}

The previous sections looked at the application of the PCA/clustering methodology to the simulation results of the experimental test rig operating at low power. This section will discuss the application of the methodology to the detailed single sector combustor operating at high power, which is a more realistic representation of the aeroengine combustor flow field and operating conditions (see \cref{sec:investigated-geometries}).

\begin{figure}[h]
	\begin{subfigure}{.5\textwidth}
		\includegraphics[trim={15cm, 0cm, 16cm, 2cm}, clip, width=0.8\linewidth]{figs/RR_combustor_slice_labelled_hidden_internals.pdf}
		\captionsetup{width=.9\linewidth}
		\caption{Combustor mid-plane with indication of all oxidiser inlets}
		\label{subfig:detailed-rql-oxidisers}
	\end{subfigure}
	\begin{subfigure}{.5\textwidth}
		\includegraphics[trim={3.5cm, 0.5cm, 2cm, 2cm}, clip, width=1.1\linewidth]{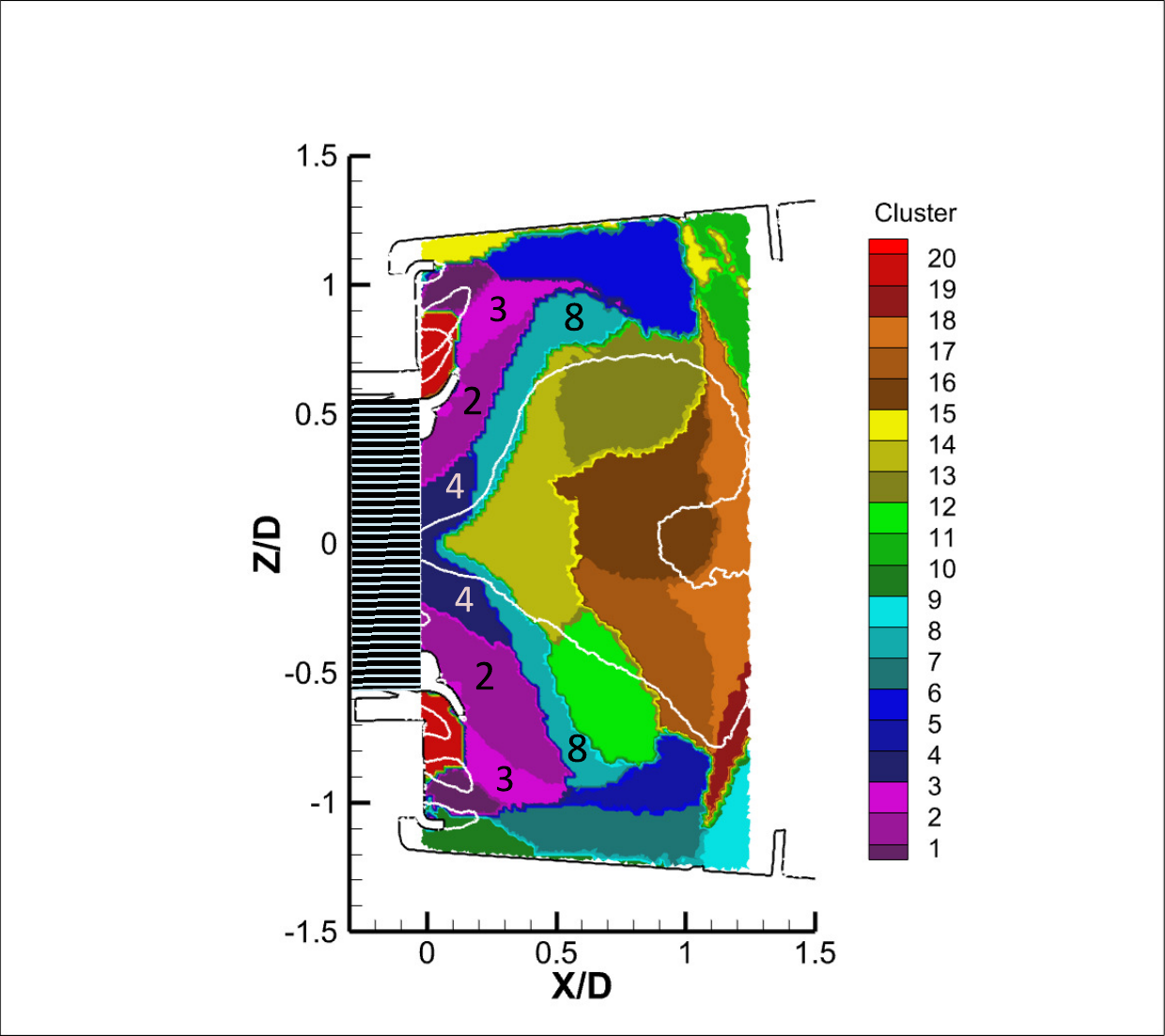}
		\captionsetup{width=.9\linewidth}
		\caption{Converged solution of the algorithm with $K=20$ clusters}
		\label{subfig:detailed-rql-mid-plane-clustering}
	\end{subfigure}
\caption{Mid-plane of the single-sector combustor showing all oxidizer inlets into the chamber and the converged solution of the K-means clustering algorithm with 20 clusters.
The white isoline is the zero mean axial velocity level ($\overline{u}_x = 0$).
The coordinates have been normalized by one injector diameter $D$.
The internal components of the fuel injector have been concealed due to commercial confidentiality.}
\label{fig:detailed-single-sector-clustering}
\end{figure}

\cref{fig:detailed-single-sector-clustering} shows the converged solution of the clustering algorithm applied to the combustor mid-plane. This dataset consists of approximately 10,000 uniformly distributed spatial locations, with a matrix of 10 passive scalars and approximately 150 temporal samples at each spatial location. Algorithm convergence was achieved in approximately 5 minutes on a 64 GB RAM desktop workstation using MATLAB R2021a. The $S_{PCA}$ weighting factor $\alpha$ (\cref{equation:combined-similarity-factor}) was set to 0.67, and the oxidizer passive scalars and the fuel mixture fraction were Pareto-scaled in a preprocessing step. Unlike in the experimental test rig case, which is circumferentially symmetrical about the injector center-line, the inner and outer wall film cooling flows in the single sector combustor are tagged by two separate passive scalars. Following the analysis described in \cref{sec:choosing-number-clusters}, the number of clusters was chosen to be $K=20$, which gives a good balance between the discretization of the dataset and the total number of clusters to be visually analyzed.

\begin{figure}
	\begin{subfigure}{.5\textwidth}
		\includegraphics[width=1\linewidth]{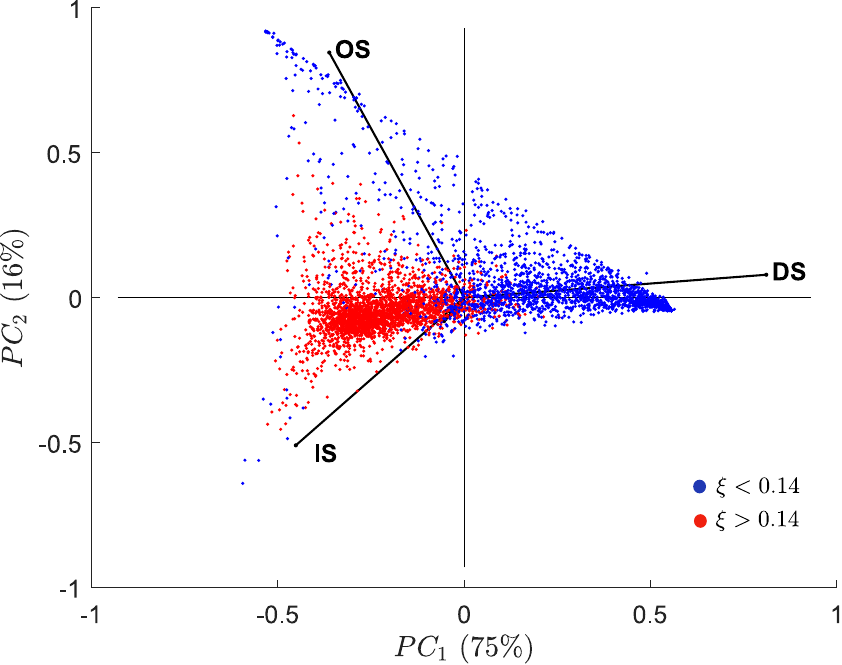}
		\caption{Cluster 2 ($\overline{\xi} = 0.14$)}
		\label{subfig:biplot-cluster-2-detailed-rql}
	\end{subfigure}
	\begin{subfigure}{.5\textwidth}
		\includegraphics[width=1\linewidth]{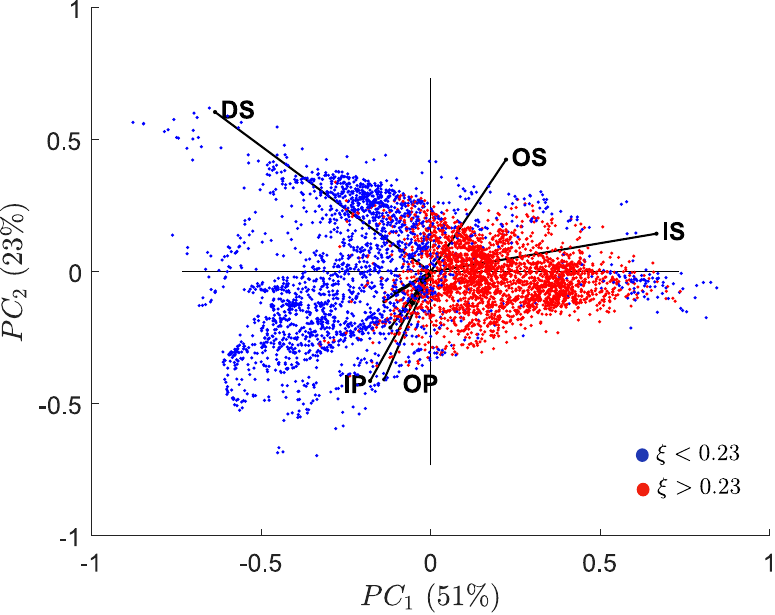}
		\caption{Cluster 8 ($\overline{\xi} = 0.23$)}
		\label{subfig:biplot-cluster-8-detailed-rql}
	\end{subfigure}
	\vspace{1cm}
	\newline
	\begin{subfigure}{.5\textwidth}
		\centering
		\includegraphics[width=1\linewidth]{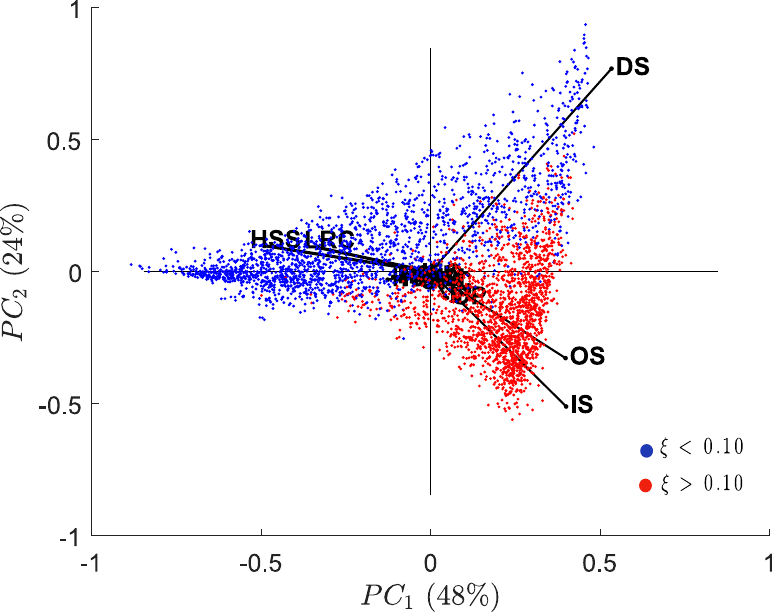}
		\caption{Cluster 3 ($\overline{\xi} = 0.10$)}
		\label{subfig:biplot-cluster-3-detailed-rql}
	\end{subfigure}
	\begin{subfigure}{.5\textwidth}
	\centering
		\includegraphics[trim={0, 0, 1cm, 0}, clip, width=1\linewidth]{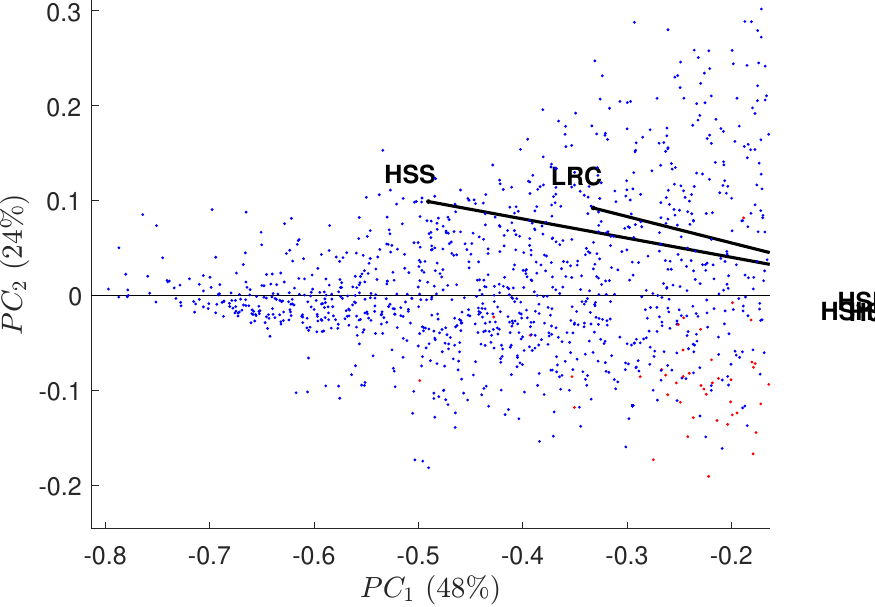}
		\caption{Close-up view of cluster 3}
		\label{subfig:biplot-cluster-3-detailed-rql-closeup}
	\end{subfigure}
\caption{Biplots of clusters 2, 3 and 8 of the detailed RQL combustor mid-plane.
The percentage of total variance explained by each PC, as well as the mean time-averaged mixture fraction across each cluster, is indicated in brackets.}
\label{fig:biplot-clusters-2-3-8-detailed-rql}
\end{figure}

Similarly to the analysis of the experimental test rig in the previous sections, the converged solution of \cref{subfig:detailed-rql-mid-plane-clustering} shows considerable symmetry in the cluster arrangement about the injector center-line, with major discrepancies between the top and bottom halves resulting from the labeling of the inner and outer film cooling flows, as well as the inner and outer port flows, with separate passive scalars. Cluster 4 is very close to the fuel prefilmer, where the liquid fuel is injected into the chamber, and the cluster is therefore dominated by the liquid fuel spray, which has not yet completed evaporation. \cref{fig:biplot-clusters-2-3-8-detailed-rql} shows the biplots across clusters 2, 3 and 8 of \cref{subfig:detailed-rql-mid-plane-clustering}. The biplots were produced following the strategy described in \cref{sec:biplots}, i.e. performing PCA on the Pareto-scaled oxidizer streams and using the mixture fraction to color the observations projected onto the space of the first two PCs. Cluster 2 is clearly dominated by the inner swirler (IS), the outer swirler (OS) and the dome swirler (DS) flows. The IS is pointing in the direction of the fuel-rich data cloud, indicating that a significant amount of fuel has evaporated from the liquid spray and has had time to mix with the inner swirler flow by the time the fuel reaches cluster 2. \cref{subfig:biplot-cluster-2-detailed-rql} also shows that the IS and the DS are strongly anti-correlated, indicating that they compete against each other over time in the mixing across cluster 2. The OS, on the other hand, is not strongly linearly correlated to either the IS or the DS, indicating that some amount of mixing may have occurred (especially with the IS), but that this mixing is far from complete, which seems reasonable given the proximity of cluster 2 to the injector outlet.

\newpage

\cref{subfig:biplot-cluster-8-detailed-rql} shows that the correlation between the IS and the OS flows across cluster 8 is considerably increased compared to cluster 2, indicating a relatively high degree of mixing between the two streams by the time they reach this cluster. The DS is still the dominant oxidiser contributing toward leaning the mixture, but the Inner Port (IP) and Outer Port (OP) flows also have a non-negligible role. \cref{subfig:biplot-cluster-3-detailed-rql} shows that in cluster 3, the OS and IS are strongly positively correlated with each other and with the fuel mixture fraction (i.e. they are the main fuel carriers into cluster 3). Unlike in clusters 2 and 8, the DS is no longer strongly anti-correlated to the IS, indicating that some degree of mixing between the two streams has taken place. Due to the proximity of cluster 3 to their respective outlets, the Heat Shield cooling (HSS) and Locating Ring Cooling (LRC) streams have a relevant role in the mixing across this cluster. They have large magnitude vectors that clearly point in the direction of the fuel-lean data cloud, and are strongly anti-correlated with the IS and OS, indicating they are the main oxidizers contributing toward leaning the mixture across this cluster.

\begin{figure}[h]
	\begin{subfigure}{.4\textwidth}
		\includegraphics[trim={3cm, 0.4cm, 5.8cm, 2cm}, clip, width=1\linewidth]{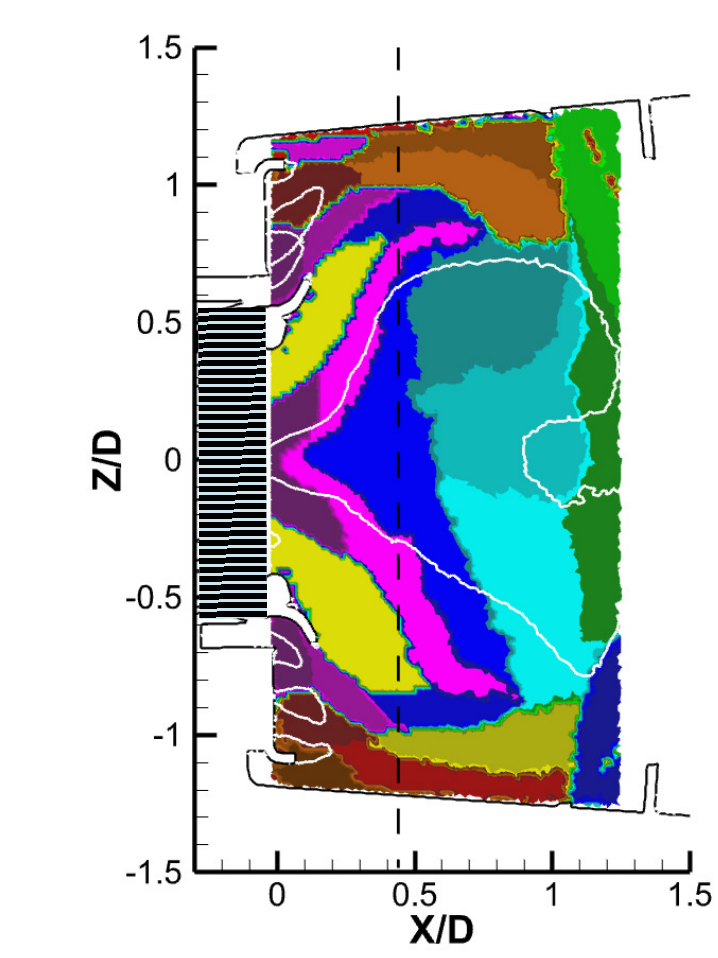}
		\caption{XZ plane}
		\label{subfig:detailed-rql-mid-plane-clustering-xz-plane}
	\end{subfigure}
	\begin{subfigure}{.6\textwidth}
		\includegraphics[trim={0.15cm, 0.2cm, 2cm, 2cm}, clip, width=1\linewidth]{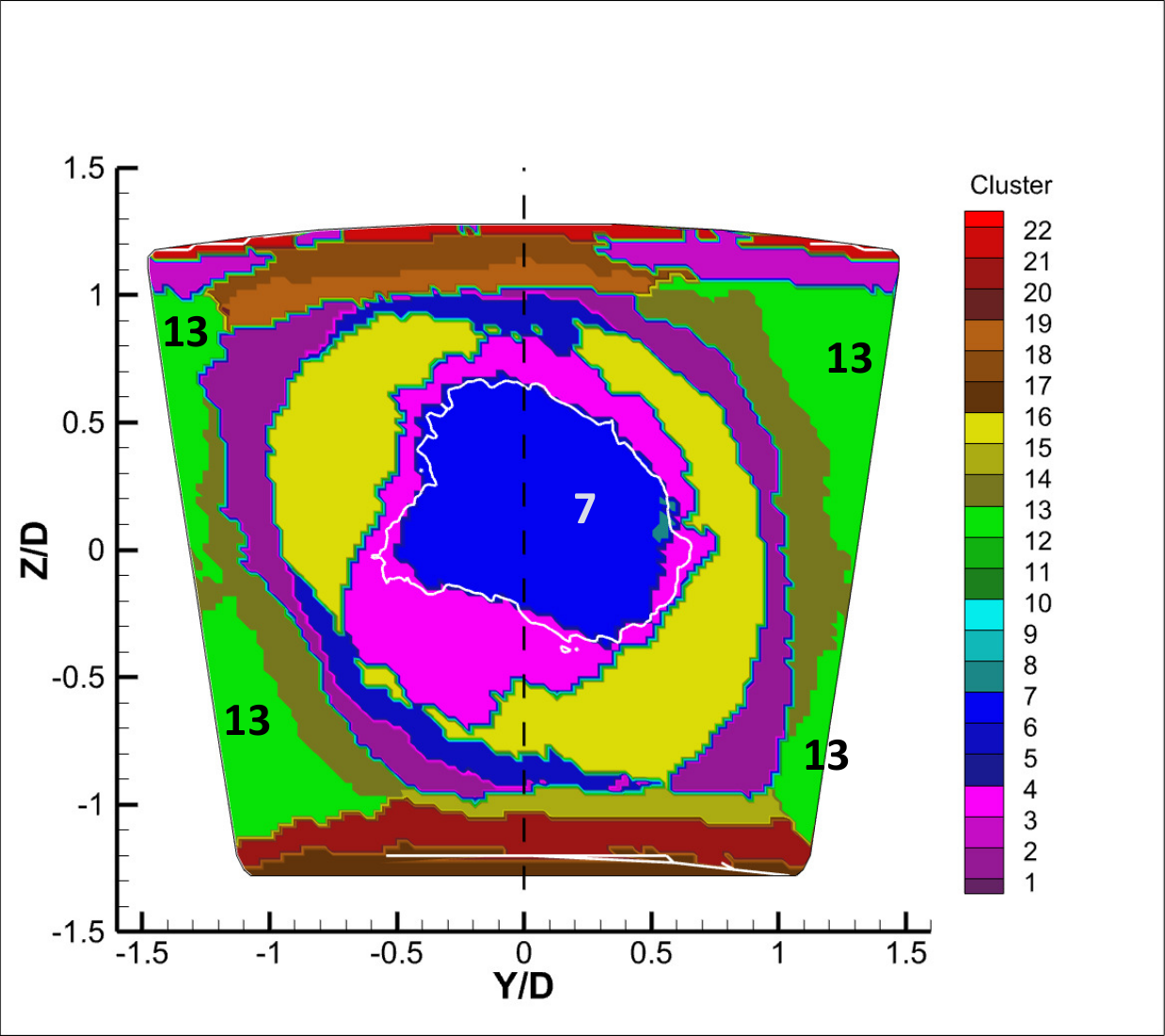}
		\caption{YZ plane, looking into the injector from downstream}
		\label{subfig:detailed-rql-mid-plane-clustering-yz-plane}
	\end{subfigure}
\caption{Converged solution of the clustering algorithm with $K=22$ clusters for a dataset of the single-sector combustor consisting of two perpendicular planes.
The dashed black line indicates the location of the perpendicular plane.
The white isoline is the zero mean axial velocity level ($\overline{u}_x = 0$).
The coordinates have been normalized by one injector diameter $D$.
The internal components of the fuel injector have been concealed due to commercial confidentiality.}

\label{fig:detailed-rql-two-planes-clustering}
\end{figure}

Finally, \cref{fig:detailed-rql-two-planes-clustering} shows the converged solution of the clustering algorithm (with a weighting factor $\alpha=0.67$) applied to a dataset consisting of two perpendicular planes of the single-sector combustor, illustrating that the algorithm can be applied to any spatial distribution. With the additional plane, the dataset becomes significantly large: approximately 19,000 spatial locations. Due to the increased dataset size, the number of clusters was increased to $K=22$. In \cref{subfig:detailed-rql-mid-plane-clustering-yz-plane}, it is interesting to point out that cluster 7 matches the shape of the zero-axial-velocity line, and the continuity of cluster 13 across the side walls due to the imposed periodic boundary condition in the LES calculation.

\FloatBarrier

\section{Conclusion}

This work presents a newly developed and computationally inexpensive statistical method for the analysis of fuel-air mixing in gas turbine combustors. It is applied as a post-processing step to data sampled from a reacting-flow LES calculation where every oxidizer and fuel inflow into the chamber is tagged with a unique passive scalar that allows it to be traced across space and time. After appropriate data scaling, PCA is employed to produce a low-order model of the passive scalar times series at each sampled spatial location. The low-order model summarizes the correlation structure of the data and can be visualized in a two-dimensional biplot, which gives insight into the local mixing of the fuel stream with the tens of oxidizer inflows that typically compose a modern gas turbine combustor, allowing the influence of individual oxidizer streams on the mixing to be identified. In order to find spatial locations across the combustor with similar mixing characteristics, a modified version of the K-means clustering algorithm, which employs a combination of PCA similarity and Mahalanobis distance as the distance metric, is applied. The clusters of spatial locations identified by the algorithm agree well with the flow physics inside a gas turbine combustor, and allow to employ a single PCA biplot to visualize the fuel-air mixing across a large number of spatial locations. Since fuel-air mixing is crucial to all combustor aerothermal processes, the proposed statistical methodology aims to help engineers better understand its complexity in realistic chambers and make better informed design decisions. Future work should investigate improved data preprocessing strategies (e.g. data scaling), and the potential of non-linear dimensionality reduction methods to better represent fuel-air mixing, although this is likely to come at the expense of increased computational complexity and reduced interpretability.

\clearpage

\bibliography{sn-article.bib}


\begin{thebibliography}{25}
\ifx \bisbn   \undefined \def \bisbn  #1{ISBN #1}\fi
\ifx \binits  \undefined \def \binits#1{#1}\fi
\ifx \bauthor  \undefined \def \bauthor#1{#1}\fi
\ifx \batitle  \undefined \def \batitle#1{#1}\fi
\ifx \bjtitle  \undefined \def \bjtitle#1{#1}\fi
\ifx \bvolume  \undefined \def \bvolume#1{\textbf{#1}}\fi
\ifx \byear  \undefined \def \byear#1{#1}\fi
\ifx \bissue  \undefined \def \bissue#1{#1}\fi
\ifx \bfpage  \undefined \def \bfpage#1{#1}\fi
\ifx \blpage  \undefined \def \blpage #1{#1}\fi
\ifx \burl  \undefined \def \burl#1{\textsf{#1}}\fi
\ifx \doiurl  \undefined \def \doiurl#1{\url{https://doi.org/#1}}\fi
\ifx \betal  \undefined \def \betal{\textit{et al.}}\fi
\ifx \binstitute  \undefined \def \binstitute#1{#1}\fi
\ifx \binstitutionaled  \undefined \def \binstitutionaled#1{#1}\fi
\ifx \bctitle  \undefined \def \bctitle#1{#1}\fi
\ifx \beditor  \undefined \def \beditor#1{#1}\fi
\ifx \bpublisher  \undefined \def \bpublisher#1{#1}\fi
\ifx \bbtitle  \undefined \def \bbtitle#1{#1}\fi
\ifx \bedition  \undefined \def \bedition#1{#1}\fi
\ifx \bseriesno  \undefined \def \bseriesno#1{#1}\fi
\ifx \blocation  \undefined \def \blocation#1{#1}\fi
\ifx \bsertitle  \undefined \def \bsertitle#1{#1}\fi
\ifx \bsnm \undefined \def \bsnm#1{#1}\fi
\ifx \bsuffix \undefined \def \bsuffix#1{#1}\fi
\ifx \bparticle \undefined \def \bparticle#1{#1}\fi
\ifx \barticle \undefined \def \barticle#1{#1}\fi
\bibcommenthead
\ifx \bconfdate \undefined \def \bconfdate #1{#1}\fi
\ifx \botherref \undefined \def \botherref #1{#1}\fi
\ifx \url \undefined \def \url#1{\textsf{#1}}\fi
\ifx \bchapter \undefined \def \bchapter#1{#1}\fi
\ifx \bbook \undefined \def \bbook#1{#1}\fi
\ifx \bcomment \undefined \def \bcomment#1{#1}\fi
\ifx \oauthor \undefined \def \oauthor#1{#1}\fi
\ifx \citeauthoryear \undefined \def \citeauthoryear#1{#1}\fi
\ifx \endbibitem  \undefined \def \endbibitem {}\fi
\ifx \bconflocation  \undefined \def \bconflocation#1{#1}\fi
\ifx \arxivurl  \undefined \def \arxivurl#1{\textsf{#1}}\fi
\csname PreBibitemsHook\endcsname

\bibitem[\protect\citeauthoryear{Lefebvre and Ballal}{2010}]{Lefebvre2010}
\begin{bbook}
\bauthor{\bsnm{Lefebvre}, \binits{A.H.}},
\bauthor{\bsnm{Ballal}, \binits{D.R.}}:
\bbtitle{{Gas Turbine Combustion: Alternative Fuels and Emissions}},
\bedition{3}rd edn.
\bpublisher{CRC Press},
\blocation{Boca Raton}
(\byear{2010})
\end{bbook}
\endbibitem

\bibitem[\protect\citeauthoryear{Lefebvre and McDonell}{2017}]{Lefebvre2017}
\begin{bbook}
\bauthor{\bsnm{Lefebvre}, \binits{A.H.}},
\bauthor{\bsnm{McDonell}, \binits{V.G.}}:
\bbtitle{{Atomization and Sprays}},
\bedition{2}nd edn.
\bpublisher{CRC Press},
\blocation{Boca Raton}
(\byear{2017})
\end{bbook}
\endbibitem

\bibitem[\protect\citeauthoryear{{Rolls-Royce}}{2015}]{Rolls-Royce2015}
\begin{bbook}
\bauthor{\bsnm{{Rolls-Royce}}}:
\bbtitle{{The Jet Engine}},
\bedition{5}th edn.
\bpublisher{Wiley},
\blocation{UK}
(\byear{2015})
\end{bbook}
\endbibitem

\bibitem[\protect\citeauthoryear{Brend et~al.}{2020}]{Brend2020}
\begin{botherref}
\oauthor{\bsnm{Brend}, \binits{M.A.}},
\oauthor{\bsnm{Denman}, \binits{P.A.}},
\oauthor{\bsnm{Carrotte}, \binits{J.F.}}:
{Volumetric PIV measurement for capturing the port flow characteristics within annular gas turbine combustors}.
Experiments in Fluids
\textbf{61}
(2020)
\doiurl{10.1007/s00348-020-2938-4}
\end{botherref}
\endbibitem

\bibitem[\protect\citeauthoryear{Hughes and Carrotte}{2004}]{Hughes2004}
\begin{barticle}
\bauthor{\bsnm{Hughes}, \binits{N.J.}},
\bauthor{\bsnm{Carrotte}, \binits{J.F.}}:
\batitle{{Unsteadiness of the port feed and jet flows within a gas turbine combustion system}}.
\bjtitle{Journal of Fluids Engineering, Transactions of the ASME}
\bvolume{126}(\bissue{1}),
\bfpage{55}--\blpage{62}
(\byear{2004})
\doiurl{10.1115/1.1637629}
\end{barticle}
\endbibitem

\bibitem[\protect\citeauthoryear{Jolliffe}{2002}]{Jolliffe2002}
\begin{bbook}
\bauthor{\bsnm{Jolliffe}, \binits{I.T.}}:
\bbtitle{{Principal Component Analysis}},
\bedition{2}nd edn.
\bsertitle{Springer Series in Statistics}.
\bpublisher{Springer},
\blocation{New York}
(\byear{2002})
\end{bbook}
\endbibitem

\bibitem[\protect\citeauthoryear{Parente et~al.}{2009}]{Parente2009}
\begin{barticle}
\bauthor{\bsnm{Parente}, \binits{A.}},
\bauthor{\bsnm{Sutherland}, \binits{J.C.}},
\bauthor{\bsnm{Tognotti}, \binits{L.}},
\bauthor{\bsnm{Smith}, \binits{P.J.}}:
\batitle{{Identification of low-dimensional manifolds in turbulent flames}}.
\bjtitle{Proc. Combust. Inst.}
\bvolume{32}(\bissue{1}),
\bfpage{1579}--\blpage{1586}
(\byear{2009})
\doiurl{10.1016/j.proci.2008.06.177}
\end{barticle}
\endbibitem

\bibitem[\protect\citeauthoryear{D'Alessio et~al.}{2020}]{DAlessio2020}
\begin{bchapter}
\bauthor{\bsnm{D'Alessio}, \binits{G.}},
\bauthor{\bsnm{Attili}, \binits{A.}},
\bauthor{\bsnm{Cuoci}, \binits{A.}},
\bauthor{\bsnm{Pitsch}, \binits{H.}},
\bauthor{\bsnm{Parente}, \binits{A.}}:
\bctitle{Analysis of turbulent reacting jets via principal component analysis}.
In: \beditor{\bsnm{Pitsch}, \binits{H.}},
\beditor{\bsnm{Attili}, \binits{A.}} (eds.)
\bbtitle{Data Analysis for Direct Numerical Simulations of Turbulent Combustion},
pp. \bfpage{233}--\blpage{251}.
\bpublisher{Springer},
\blocation{Cham}
(\byear{2020}).
\doiurl{10.1007/978-3-030-44718-2_12}
\end{bchapter}
\endbibitem

\bibitem[\protect\citeauthoryear{Berkooz et~al.}{1993}]{Berkooz1993}
\begin{barticle}
\bauthor{\bsnm{Berkooz}, \binits{G.}},
\bauthor{\bsnm{Holmes}, \binits{P.}},
\bauthor{\bsnm{Lumley}, \binits{J.L.}}:
\batitle{{The Proper Orthogonal Decomposition in the Analysis of Turbulent Flows}}.
\bjtitle{Annual Review of Fluid Mechanics}
\bvolume{25}(\bissue{1}),
\bfpage{539}--\blpage{575}
(\byear{1993})
\doiurl{10.1146/annurev.fl.25.010193.002543}
\end{barticle}
\endbibitem

\bibitem[\protect\citeauthoryear{Brunton and Kutz}{2019}]{Brunton2019}
\begin{bbook}
\bauthor{\bsnm{Brunton}, \binits{S.L.}},
\bauthor{\bsnm{Kutz}, \binits{J.N.}}:
\bbtitle{Data-Driven Science and Engineering: Machine Learning, Dynamical Systems, and Control}.
\bpublisher{Cambridge University Press},
\blocation{Cambridge}
(\byear{2019}).
\doiurl{10.1017/9781108380690}
\end{bbook}
\endbibitem

\bibitem[\protect\citeauthoryear{Poinsot and Veynante}{2005}]{Poinsot}
\begin{bbook}
\bauthor{\bsnm{Poinsot}, \binits{T.}},
\bauthor{\bsnm{Veynante}, \binits{D.}}:
\bbtitle{{Theoretical and Numerical Combustion}},
\bedition{3rd} edn.
\bpublisher{RT Edwards, Inc.},
\blocation{Philadelphia}
(\byear{2005})
\end{bbook}
\endbibitem

\bibitem[\protect\citeauthoryear{Chen et~al.}{2020}]{Chen2020}
\begin{bchapter}
\bauthor{\bsnm{Chen}, \binits{Z.X.}},
\bauthor{\bsnm{Langella}, \binits{I.}},
\bauthor{\bsnm{Swaminathan}, \binits{N.}}:
\bctitle{{The Role of CFD in Modern Jet Engine Combustor Design}}.
In: \beditor{\bsnm{Agarwal}, \binits{R.K.}} (ed.)
\bbtitle{Environmental Impact of Aviation and Sustainable Solutions}.
\bpublisher{IntechOpen},
\blocation{London}
(\byear{2020}).
\doiurl{10.5772/intechopen.88267}
\end{bchapter}
\endbibitem

\bibitem[\protect\citeauthoryear{Veynante and Vervisch}{2002}]{Veynante2002}
\begin{barticle}
\bauthor{\bsnm{Veynante}, \binits{D.}},
\bauthor{\bsnm{Vervisch}, \binits{L.}}:
\batitle{Turbulent combustion modeling}.
\bjtitle{Progress in Energy and Combustion Science}
\bvolume{28}(\bissue{3}),
\bfpage{193}--\blpage{266}
(\byear{2002})
\doiurl{10.1016/S0360-1285(01)00017-X}
\end{barticle}
\endbibitem

\bibitem[\protect\citeauthoryear{Vervisch and Roekaerts}{2015}]{Vervisch2015}
\begin{bbook}
\beditor{\bsnm{Vervisch}, \binits{L.}},
\beditor{\bsnm{Roekaerts}, \binits{D.}} (eds.):
\bbtitle{{Best Practice Guidelines for CFD of Turbulent Combustion}}.
\bpublisher{ERCOFTAC},
\blocation{UK}
(\byear{2015})
\end{bbook}
\endbibitem

\bibitem[\protect\citeauthoryear{Pope}{2000}]{Pope2000}
\begin{bbook}
\bauthor{\bsnm{Pope}, \binits{S.B.}}:
\bbtitle{{Turbulent Flows}},
\bedition{2nd} edn.
\bpublisher{Cambridge University Press},
\blocation{Cambridge}
(\byear{2000}).
\doiurl{10.1017/CBO9780511840531}
\end{bbook}
\endbibitem

\bibitem[\protect\citeauthoryear{Fiorina et~al.}{2015}]{Fiorina2015}
\begin{barticle}
\bauthor{\bsnm{Fiorina}, \binits{B.}},
\bauthor{\bsnm{Veynante}, \binits{D.}},
\bauthor{\bsnm{Candel}, \binits{S.}}:
\batitle{{Modeling Combustion Chemistry in Large Eddy Simulation of Turbulent Flames}}.
\bjtitle{Flow, Turbulence and Combustion}
\bvolume{94}(\bissue{1}),
\bfpage{3}--\blpage{42}
(\byear{2015})
\doiurl{10.1007/s10494-014-9579-8}
\end{barticle}
\endbibitem

\bibitem[\protect\citeauthoryear{Peters}{2000}]{Peters2000}
\begin{bbook}
\bauthor{\bsnm{Peters}, \binits{N.}}:
\bbtitle{{Turbulent Combustion}}.
\bpublisher{Cambridge University Press},
\blocation{Cambridge}
(\byear{2000}).
\doiurl{10.1017/CBO9780511612701}
\end{bbook}
\endbibitem

\bibitem[\protect\citeauthoryear{Giusti et~al.}{2019}]{Giusti2019}
\begin{botherref}
\oauthor{\bsnm{Giusti}, \binits{A.}},
\oauthor{\bsnm{Magri}, \binits{L.}},
\oauthor{\bsnm{Zedda}, \binits{M.}}:
{Flow Inhomogeneities in a Realistic Aeronautical Gas-Turbine Combustor: Formation, Evolution, and Indirect Noise}.
ASME. J. Eng. Gas Turbines Power
\textbf{141}(1)
(2019)
\doiurl{10.1115/1.4040810}
\end{botherref}
\endbibitem

\bibitem[\protect\citeauthoryear{Parente and Sutherland}{2013}]{Parente2013}
\begin{barticle}
\bauthor{\bsnm{Parente}, \binits{A.}},
\bauthor{\bsnm{Sutherland}, \binits{J.C.}}:
\batitle{{Principal component analysis of turbulent combustion data: Data pre-processing and manifold sensitivity}}.
\bjtitle{Combust. Flame}
\bvolume{160}(\bissue{2}),
\bfpage{340}--\blpage{350}
(\byear{2013})
\doiurl{10.1016/j.combustflame.2012.09.016}
\end{barticle}
\endbibitem

\bibitem[\protect\citeauthoryear{van~den Berg et~al.}{2006}]{VandenBerg2006}
\begin{botherref}
\oauthor{\bsnm{Berg}, \binits{R.A.}},
\oauthor{\bsnm{Hoefsloot}, \binits{H.C.J.}},
\oauthor{\bsnm{Westerhuis}, \binits{J.A.}},
\oauthor{\bsnm{Smilde}, \binits{A.K.}},
\oauthor{\bsnm{Werf}, \binits{M.J.}}:
Centering, scaling, and transformations: Improving the biological information content of metabolomics data.
BMC Genomics
\textbf{7}(1)
(2006)
\doiurl{10.1186/1471-2164-7-142}
\end{botherref}
\endbibitem

\bibitem[\protect\citeauthoryear{Jolliffe and Cadima}{2016}]{Jolliffe2016}
\begin{botherref}
\oauthor{\bsnm{Jolliffe}, \binits{I.T.}},
\oauthor{\bsnm{Cadima}, \binits{J.}}:
{Principal component analysis: a review and recent developments}.
Philosophical Transactions of the Royal Society A: Mathematical, Physical and Engineering Sciences
\textbf{374}(2065)
(2016)
\doiurl{10.1098/rsta.2015.0202}
\end{botherref}
\endbibitem

\bibitem[\protect\citeauthoryear{Varmuza and Filzmoser}{2016}]{Varmuza2016}
\begin{bbook}
\bauthor{\bsnm{Varmuza}, \binits{K.}},
\bauthor{\bsnm{Filzmoser}, \binits{P.}}:
\bbtitle{{Introduction to Multivariate Statistical Analysis in Chemometrics}}.
\bpublisher{CRC Press},
\blocation{Boca Raton}
(\byear{2016}).
\doiurl{10.1201/9781420059496}
\end{bbook}
\endbibitem

\bibitem[\protect\citeauthoryear{Singhal and Seborg}{2005}]{Singhal2005}
\begin{barticle}
\bauthor{\bsnm{Singhal}, \binits{A.}},
\bauthor{\bsnm{Seborg}, \binits{D.E.}}:
\batitle{Clustering multivariate time‐series data}.
\bjtitle{Journal of Chemometrics}
\bvolume{19}(\bissue{8}),
\bfpage{427}--\blpage{438}
(\byear{2005})
\doiurl{10.1002/cem.945}
\end{barticle}
\endbibitem

\bibitem[\protect\citeauthoryear{Johannesmeyer}{1999}]{Johannesmeyer1999}
\begin{botherref}
\oauthor{\bsnm{Johannesmeyer}, \binits{M.C.}}:
Abnormal situation analysis using pattern recognition techniques and historical data.
Master's thesis,
University of California Santa Barbara
(1999)
\end{botherref}
\endbibitem

\bibitem[\protect\citeauthoryear{Singhal and Seborg}{2002}]{Singhal2002}
\begin{barticle}
\bauthor{\bsnm{Singhal}, \binits{A.}},
\bauthor{\bsnm{Seborg}, \binits{D.E.}}:
\batitle{{Pattern Matching in Multivariate Time Series Databases Using a Moving-Window Approach}}.
\bjtitle{Industrial \& Engineering Chemistry Research}
\bvolume{41}(\bissue{16}),
\bfpage{3822}--\blpage{3838}
(\byear{2002})
\doiurl{10.1021/ie010517z}
\end{barticle}
\endbibitem

\end{thebibliography}

\end{document}